\documentclass[12pt,preprint]{aastex}
\usepackage{emulateapj5,apjfonts}
\usepackage{graphics}
\usepackage{amssymb}
\usepackage{bbm}
\usepackage{amsmath,textcomp,array}
\usepackage{onecolfloat}
\usepackage{bm}

\lefthead{Heitmann et al.}
\righthead{Cosmological Models and Precision 
Emulation of the Nonlinear Matter Power Spectrum}

\begin{document}


\def\head{
  \vbox to 0pt{\vss
                    \hbox to 0pt{\hskip 440pt\rm LA-UR-08-07921\hss}
                   \vskip 25pt}

\title{The Coyote Universe II: Cosmological Models and Precision
Emulation of the Nonlinear Matter Power Spectrum}
\author{Katrin~Heitmann\altaffilmark{1}, David Higdon\altaffilmark{2},
Martin White\altaffilmark{3}, Salman~Habib\altaffilmark{4}, Brian
J. Williams\altaffilmark{2}, Earl Lawrence\altaffilmark{2},  and Christian Wagner\altaffilmark{5}}

\affil{$^1$ ISR-1, ISR Division,  Los
Alamos National Laboratory, Los Alamos, NM 87545}
\affil{$^2$ CCS-6, CCS Division, Los Alamos National Laboratory, Los Alamos, NM 87545}
\affil{$^3$ Departments of Physics and Astronomy, University of
California, Berkeley, CA 94720}
\affil{$^4$ T-2, Theoretical Division, Los
Alamos National Laboratory, Los Alamos, NM 87545}
\affil{$^5$ Astrophysikalisches Institut Potsdam (AIP),
An der Sternwarte 16, D-14482 Potsdam}

\date{today}

\begin{abstract}
  The power spectrum of density fluctuations is a foundational source
  of cosmological information. Precision cosmological probes targeted
  primarily at investigations of dark energy require accurate
  theoretical determinations of the power spectrum in the nonlinear
  regime. To exploit the observational power of future cosmological
  surveys, accuracy demands on the theory are at the one percent level
  or better. Numerical simulations are currently the only way to
  produce sufficiently error-controlled predictions for the power
  spectrum. The very high computational cost of (precision) $N$-body
  simulations is a major obstacle to obtaining predictions in the
  nonlinear regime, while scanning over cosmological
  parameters. Near-future observations, however, are likely to provide
  a meaningful constraint only on constant dark energy equation of
  state, `$w$CDM', cosmologies. In this paper we demonstrate that a
  limited set of only 37 cosmological models -- the ``Coyote
  Universe'' suite -- can be used to predict the nonlinear matter
  power spectrum to one percent over a prior parameter range
  set by current cosmic microwave background observations. This paper is the
  second in a series of three, with the final aim to provide a
  high-accuracy prediction scheme for the nonlinear matter power
  spectrum for $w$CDM cosmologies.
\end{abstract}

\keywords{methods: statistical ---
          cosmology: large-scale structure of the universe}}

\twocolumn[\head]
\section{Introduction}

Although the discovery of cosmic acceleration by \citet{riess} and
\citet{perlmutter} is already a decade in the past, our understanding
of the nature of the underlying driver of the acceleration, ``dark
energy'', has made little progress. One reason for this is that the
dark energy equation of state parameter $w$ is consistent with a
cosmological constant ($w=-1$) at roughly 10\% accuracy, with no
constraints on any possible time dependence. In order to advance
further in terms of distinguishing different models of dark energy
from each other and dark energy itself from other possible causes of
acceleration (such as a possible break-down of general relativity on
very large scales) observational errors must be brought down
significantly. The current target is to achieve another order of
magnitude improvement for several dark energy probes -- probes that
measure not only the expansion history of the Universe but also the
growth of cosmological structure -- down to the level of a percent.

To date, the five most promising lines of investigation are:
(i) Supernovae Type Ia, to measure the expansion history of the Universe,
(ii) clusters of galaxies, to measure the expansion history and growth
of structure,
(iii) baryon acoustic oscillations, to measure the expansion history,
(iv) weak lensing, to measure the expansion history and the growth of
structure and
(v) redshift distortions to measure the growth of structure.
The last three probes, baryon acoustic oscillations, weak lensing and redshift
space distortions, rely the most strongly on precise predictions of the
nonlinear matter power spectrum.
Numerical simulations are essential for carrying out this task, not only
for the power spectrum itself, but also to build the underlying skeleton
of cosmological structure from which object catalogs can be constructed.
The resulting `mock catalogs' have many uses: to design and test observational
strategies, to understand systematic errors therein, and to confront 
theoretical predictions with observations.

In the case of baryon acoustic oscillations, measurements are carried
out on very large scales, where the nonlinear effects are
small. Therefore, higher order perturbation theory might offer an
alternative path to obtaining precise predictions for the nonlinear
matter power spectrum~(see, e.g.,
\citealt{crocce06,matsubara08,pietroni08,Car09,PadWhi09} and references
therein), and provide a useful foil for the numerical results. Weak
lensing measurements go down to much smaller spatial scales, out to 
wavenumbers $k\sim 1-10~h$Mpc$^{-1}$ (and even larger wavenumbers 
in the future). On these smaller length scales, perturbative techniques fail,
and one must rely on numerical simulations to obtain the required
level of accuracy: at least as accurate as the observations, and to be
optimal, substantially more accurate. As shown by,
e.g. \citet{hutak03}, to avoid biasing of cosmological parameter
estimations a wide-field weak lensing survey such as the
SNAP\footnote{http://snap.lbl.gov}
design requires 1\% accurate power spectrum predictions, and a survey
such as the Large Synoptic Survey Telescope
(LSST\footnote{http://www.lsst.org}) requires predictions at the
0.5\% accuracy level.

These requirements pose two major challenges: First, one must show
that simulations capturing the essential physics have reached the
desired level of accuracy. For baryon acoustic oscillations, it is
expected that gravity-only $N$-body simulations, augmented by halo
occupancy modeling, are sufficient for the task. In the case of weak
lensing, this assumption holds for scales out to $k\sim 1\,h\,$Mpc$^{-1}$.
In the first paper of this series \citep{Heitmann08} we have established
that, up to these scales, the nonlinear matter power spectrum can be
determined at sub-percent accuracy by gravity-only $N$-body simulations.
At smaller scales, baryonic physics becomes important at the few to ten
percent level and has to be taken into account
\citep{white04,zhanknox,jing06,rudd,GuiTeyCol09}, a task which has to
be tackled accurately in the near future, perhaps by a suitable
combination of simulations and observations.

After overcoming the first challenge, the next task in constraining
cosmological parameters, is to cover a range of different
cosmologies. Markov Chain Monte Carlo (MCMC) methods, commonly used for
parameter determination, rely on results from model evaluations
numbering in the tens of thousands to hundreds of thousands. Since an
accurate $N$-body simulation on the scales of interest mentioned above
costs of the order of $\sim 20,000$~CPU-hours, it is not feasible to
run such simulations for each model. (Running $\sim 20,000$ $N$-body
simulations with the required resolution on a contemporary 2048
processor cluster would take 20 years!) Taking into account the fact
that adding gasdynamics and feedback effects substantially increases
both the computational load and the number of parameters to be varied,
it is clear that a brute force approach to the problem has to be
avoided. What we need is a generalized interpolation method capable of
yielding very accurate predictions for the nonlinear matter power
spectrum from a restricted number of simulations. In the following, we
will refer to such a prediction scheme as an emulator. The emulator
will be tasked with replacing brute force $N$-body simulations for the
nonlinear matter power spectrum over a pre-defined set of cosmological
parameters, with specified ranges for the chosen parameters.

In the cosmic microwave background (CMB) community several different
paths have been suggested to provide such an emulator for the CMB
temperature anisotropy power spectrum. These include purely analytic
fits \citep{teg00,Jim04} and combinations of analytic and
semi-analytic fits \citep{kap02}. More recently, neural network
methods and machine learning techniques have been successfully used to
generate very accurate temperature anisotropy power spectra (Fendt \&
Wandelt 2007a; Auld et al. 2007; Fendt \& Wandelt 2007b; Auld et
al. 2008). These methods are based on a large number of training sets,
up to several tens of thousands. (For an alternative approach requiring many fewer 
simulations, see \citealt{HHHNW}.) While this does not constitute a
problem for anisotropy power spectra -- given the speeds at which
numerical codes such as CAMB and CMBFast can be run -- the approach is
not feasible for determining the matter power spectrum, which requires
large-scale supercomputer simulations. 

As in the case for the temperature anisotropy power spectrum, several
attempts have been made to avoid costly simulations by finding good
approximations for the nonlinear matter power spectrum. These range
from more or less analytic derivations (e.g., \citealt{ham91,pd94}) to
semi-analytic fits calibrated more specifically against simulation
results (e.g., \citealt{pd96,smith03}). Unfortunately, the accuracy of
these approximations is inadequate, at best reaching the 5-10\% level
(see, e.g., \citealt{Heitmann08} for a recent comparison of precision
simulations with {\sc HaloFit}, a fitting scheme due to
\citealt{smith03}). Thus, an order of magnitude improvement is needed
to address the accuracy requirements discussed above.

Accurate emulation is needed for many observational quantities in
cosmology, power spectra being one important example. To address
this need, we have recently introduced the ``Cosmic Calibration
Framework''~\citep{HHHN,HHHNW,SKHHHN} combining sophisticated
simulation designs with Gaussian process (GP) modeling to create very
accurate emulators from a restricted set of simulations. The term
`simulation design' refers to the specific choice of parameter
settings at which to carry out the simulations. One of the main
reasons why the Cosmic Calibration Framework provides very accurate
results from only a small number of training sets is the optimization
of the simulation design strategy to work well with the chosen
interpolation scheme, the Gaussian process in this particular
case. Another useful aspect of the methodology is that it contains an
error prediction scheme, so that one can verify the consistency of the
obtained results.

In this paper we will explain and demonstrate the emulation capability
of the Cosmic Calibration Framework. With only a small number of
simulations, an emulator for the nonlinear matter power spectrum
can be constructed, matching the simulation results at the level of
1\% accuracy. We focus on the regime of spatial wavenumber
$k\lesssim 1\,h\,{\rm Mpc}^{-1}$ and a redshift range between $z=0$ and
$z=1$, covering the current space of interest for weak lensing measurements.
Such an emulator will eliminate a major source of bias in determining
cosmological parameters from weak lensing data.
In order to design, construct, and test an emulator, it is useful to carry
out the process first on a proxy for the expensive numerical simulations;
the proxy need not be very accurate but should reflect the overall behavior
of the detailed simulations; we employ {\sc HaloFit\/} in this role.

This paper is the second in a series of three communications. In the
first, we have demonstrated that it is possible to obtain nonlinear
matter power spectra at sub-percent level accuracy out to
$k\simeq 1\,h\,{\rm Mpc}^{-1}$ from simulations, having derived and
presented a set of requirements for such simulations.
The third paper of the series will present results from the complete
simulation suite based on the cosmologies presented in the current paper,
as well as the public release of a precision power spectrum emulator.
The simulation suite is named the ``Coyote Universe'' after the computer
cluster it has been carried out on.

The paper is organized as follows. In Section \ref{cal} we describe in
detail the Cosmic Calibration Framework with special emphasis on
building a nonlinear matter power spectrum emulator from a restricted
set of simulations. We explain the design strategy for generating the
training sets and discuss the emulation step, demonstrating the
emulator accuracy. Next we provide a sensitivity analysis showing how
the power spectrum varies -- in a scale-dependent manner -- as the
cosmological parameters are changed.  We present our conclusions in
Section \ref{conclusion}.

\section{The Cosmic Calibration Framework}
\label{cal}

The Cosmic Calibration Framework~\citep{HHHN,HHHNW,SKHHHN} consists of
four interlocking steps: (i) the simulation design, which determines
at what parameter settings to generate the training sets, (ii)
generation of the emulator which replaces the simulator as a predictor
of results away from the points that were used to generate the
training set, (iii) the uncertainty and sensitivity analysis
associated with the emulator, and (iv) the calibration against data
via MCMC methods to determine parameter constraints.

In the following we discuss steps (i) - (iii) in detail, with
special emphasis on generating an accurate emulator for the nonlinear
matter power spectrum.

\subsection{Sampling the Model Space} \label{sec:design}

As discussed in the Introduction, one of the major challenges in
building an accurate emulator for the nonlinear matter power spectrum
is the very high cost of individual $N$-body simulations combined
with the high dimensionality of the parameter space (which may include
cosmological, physical, and numerical modeling parameters). To sample
the model space, the number of parameters to be varied must be
specified, as well as the range of variation for each parameter. For
now, we will assume that some combination of observational knowledge
and cosmological and astrophysical modeling is sufficient to decide on
conservative choices for sampling the model space. (We will return to this
question later, in Section~\ref{obscon}.)

Following this decision, the next step is to find a method for
sampling the model space and interpolating the results therefrom,
satisfying given accuracy criteria, and using only a manageable number
of simulation design points. In several applications, space-filling
Latin hypercube (LH) designs (details below) have proven to be well
suited for the GP model-based approach~\citep{welch89,ill91} to
solving the interpolation problem. The Cosmic Calibration Framework
uses this methodology; the associated validation examples can be found
in \citet{HHHN} and \citet{HHHNW}.

In the following, we first discuss the statistical aspects of sampling
model space followed by the observational aspects, i.e., the
parameters to be varied and the corresponding choices for the range of
variation. The parameter choices and prior ranges from observations used 
here rely on the most recent CMB observations from WMAP-5~\citep{WMAP5}.

\subsubsection{Statistical Sampling Methods}

Our first aim is to find a distribution of the parameter settings --
the simulation design -- which provides optimal coverage (in a sense
to be defined below) of the parameter space, using only a limited
number of sampling points. (In the statistics literature, it is
customary to use normalized units in which all parameters range over
the interval $[0,1]$ and we will follow this usage for the most part.)
If the actual behavior of the observable as a function of the
parameters is considered to be unknown, then it is sensible to start
with a strategy that attempts to uniformly sample the space
(space-filling design). An extreme version of this is a simple,
regular hypercubic grid. The problem with using a grid-based
interpolation method is the high dimensionality of the space. Suppose
we wish to vary 5 cosmological parameters and sample each parameter
only three times -- at near-maximum, near-minimum, and in the middle
of each parameter range. Already, this would require $3^5=243$
simulations -- not a small number -- and lead to poor coverage of the
parameter space. Such a design with only 3 levels (three sampling 
points per parameter) would also only allow for estimating a quadratic 
model. If we want to go to a higher level, the number of runs will explode 
-- if we wish to keep the complete grid. If on the other hand, 
we try to reduce the number of runs by using only a fraction of the grid, 
aliasing becomes a potential problem.

The opposite extreme of pure random sampling suffers from
clustering of sample points and occurrence of large voids in the
sampling region when the number of sample points is limited. Stratified
sampling techniques combine the idea of a regular grid and random
sampling by using strata that (equally) sub-divide the sample space,
with random sampling employed within each strata. A final important
point is that the computed observable may depend on some
sub-combination of input variables (factor sparsity), in which case we
would like to have uniform coverage across the projection of the full
space onto the relevant factors. Not all uniform sampling schemes
possess this property.

The GP model interpolation scheme used here does not require a simple
grid design. Simulation designs well-suited for GP model emulators are
LH-based designs, a type of stratified sampling scheme. Latin
hypercube sampling generalizes the Latin square for two variables,
where only one sampling point can exist in each row and each column. A
Latin hypercube sample -- in arbitrary dimensions -- consists of
points stratified in each  (axis-oriented) projection. More formally, 
an LH design is an $n\times m$ matrix in which each column is a
unique random permutation of $\{1,...,n\}$. The use of LH designs in
applications where the aim is to predict the outcome of some quantity
at untried parameter settings from a restricted set of simulations was
first proposed by
\citet{mckay79}. As discussed in more detail below, like many other
interpolators, GP models rely on local information for their
interpolation strategy. Therefore, the simulation design must provide
good coverage over the whole parameter space. Space-filling LH designs
and their variants provide an effective means for achieving this.

Very often LH designs are combined with other design strategies such
as orthogonal array (OA)-based designs or are optimized in other ways,
e.g., by symmetrizing them (more details below). By intelligently melding
design strategies, different attributes of the individual sampling
strategies can be combined to lead to improved designs, and shortcomings of 
specific designs can be eliminated. As a last step, optimization
schemes are often applied to spread out the points evenly in a
projected space. One such optimization scheme is based on minimizing
the maximal distance between points in the parameter space, which will
lead to more even coverage. We provide some details about different 
optimization schemes in Appendix~\ref{appopt}. For a discussion of different design
approaches and their specific advantages see, e.g., \citet{santner03}.

We now briefly discuss two design strategies well suited to
cosmological applications in which the number of parameters is much
less than the number of simulations that can be performed. These are
optimal OA-LH design strategies and optimal symmetric LH design
strategies. The former has been used in previous work in
cosmology~\citep{HHHN,HHHNW}, while the latter will be used in this
paper to construct the design for the Coyote Universe.  For
illustration purposes, we will use a very simple, three-dimensional
case example with three parameters, $\theta_1$, $\theta_2$, $\theta_3$
and nine sampling points.

In order to create an OA-LH design, the strategy proceeds in two
steps: (i) construction of the orthogonal array and (ii) the following construction
of the orthogonal-array based Latin hypercube.  We discuss these two
steps in turn following closely the description by \citet{tang93} and
\citet{leary03}.

\subsubsubsection{OA Designs}

An orthogonal array distributes runs uniformly in certain projections
of the full parameter space.  The mathematical definition is as follows:
An $n$ by $m$ matrix ${\bf A}$ with entries from a set ${1,2,...,s}$ is called an
orthogonal array of strength $r$, size $n$, with $m$ constraints, and
$s$ levels, if each $n\times r$ submatrix of ${\bf A}$ contains all
possible $1\times r$ rows with the same frequency $\lambda$. Here
$\lambda$ is termed the index of the array, and $n=\lambda s^r$. The
array is denoted OA$(n,m,s,r)$ \citep{tang93}.

For our application, $n$ denotes the number of simulation runs to be
performed and $m$ specifies the number of parameters to be varied
(these can be cosmological parameters as well as numerical input
parameters). These choices fix the number of dimensions in the
parameter hypercube. The parameter $s$ defines the levels of
stratification for each column in the matrix $\bf{A}$. In order to
sample the parameter space well in a uniform manner, it is often not
enough to sample it well globally. For example, if two or more
parameters interact strongly with each other, it is clearly desirable
to have a good space-filling design in the subspace of these
parameters. In other words, if one projects the design down onto, e.g,
two dimensions, such a projection should have space-filling properties
in those dimensions as well. The parameter $r$, the strength of OA
designs, indicates the projections for which the LH design based on
that OA are guaranteed to be space-filling.  For example, if $r=3$,
then all 1, 2 and 3 dimensional projections will be
space-filling. Obviously, $r$ cannot be larger than $m$, the number of
parameters varied.

The strength, $r$, is usually restricted to two or three for several
reasons: (i) Fewer and fewer OA designs with appropriate run sizes are
known as the strength increases (more on this below). (ii) In most
applications, only a small number of parameters influence the response
significantly. Statisticians call this the ``20-80 rule'' -- 20\% of
the parameters being responsible for 80\% of the outcome variation.
Therefore, the aim is to capture the action of these relevant
parameters. Furthermore, outcome variation is often dominated by a
small number of single parameter and two-factor interaction effects,
which are adequately covered by OA-LH designs based on $r=2$ or 3.
(iii) The number of simulations often has to be kept small, therefore
$r$ cannot be chosen too large, since the number of simulations $n$ is
connected to $r$ via $n=\lambda s^r$. As is the case for $r$, the
stratification parameter $s$ is also restricted by the number of runs
one can possibly perform. It is very often set to $s=2$ which then
requires the number of runs to be a power of two. The frequency
$\lambda$ with which a permutation repeats, is kept small as well. To
create a design, the strategy is to fix strength first, and try to
find an OA design that has approximately the right number of runs and
at least as many parameters as one needs. If such a design cannot be
found, then the strength is reduced by one and the process repeated.
Usually, this strategy is started with OA designs of strength 3,
though many more designs of strength 2 exist. It is rarely possible to
find a strength four or higher design with few enough runs.

\begin{figure}
\centerline{
 \includegraphics[width=1.8in]{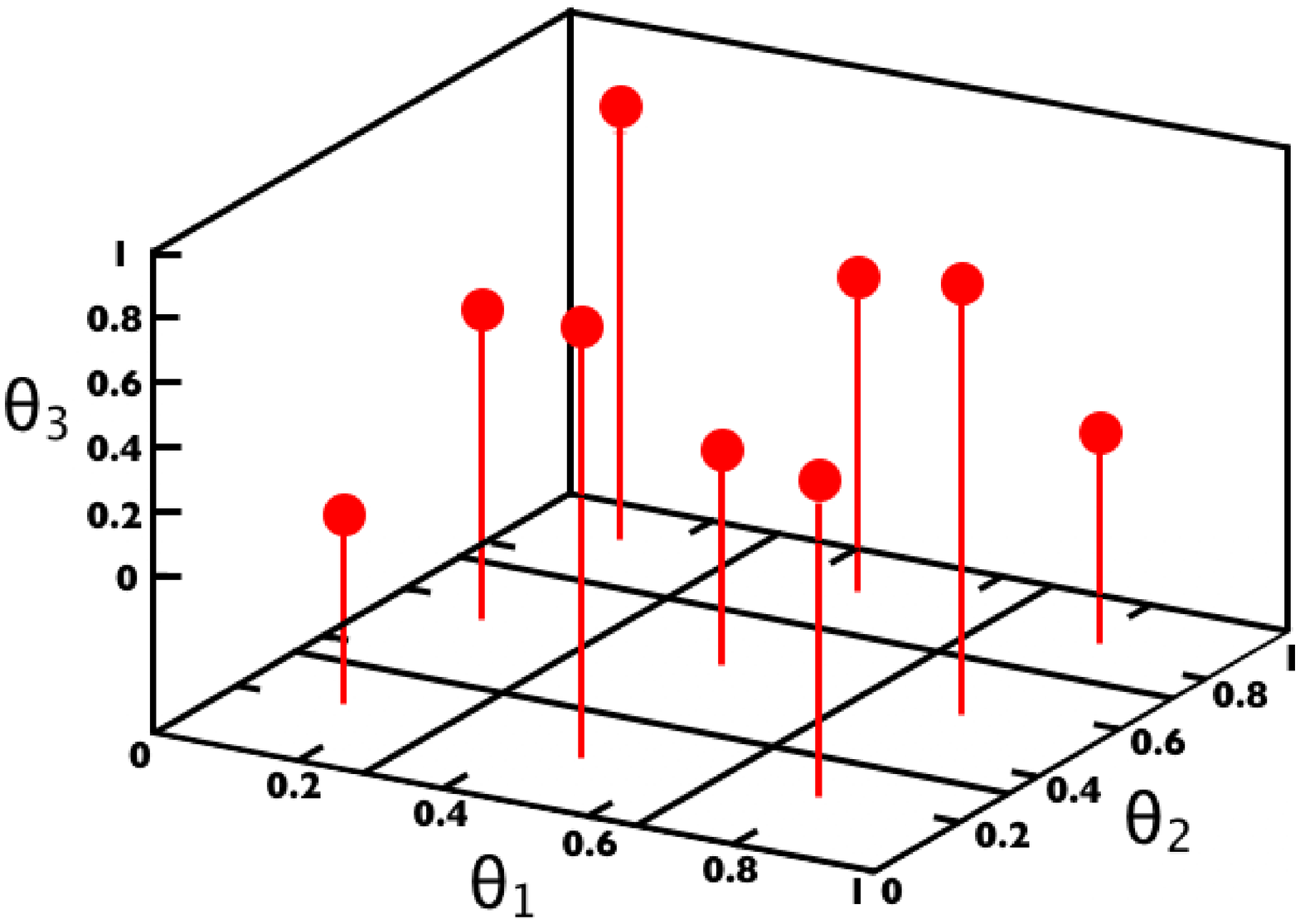}
 \includegraphics[width=1.8in]{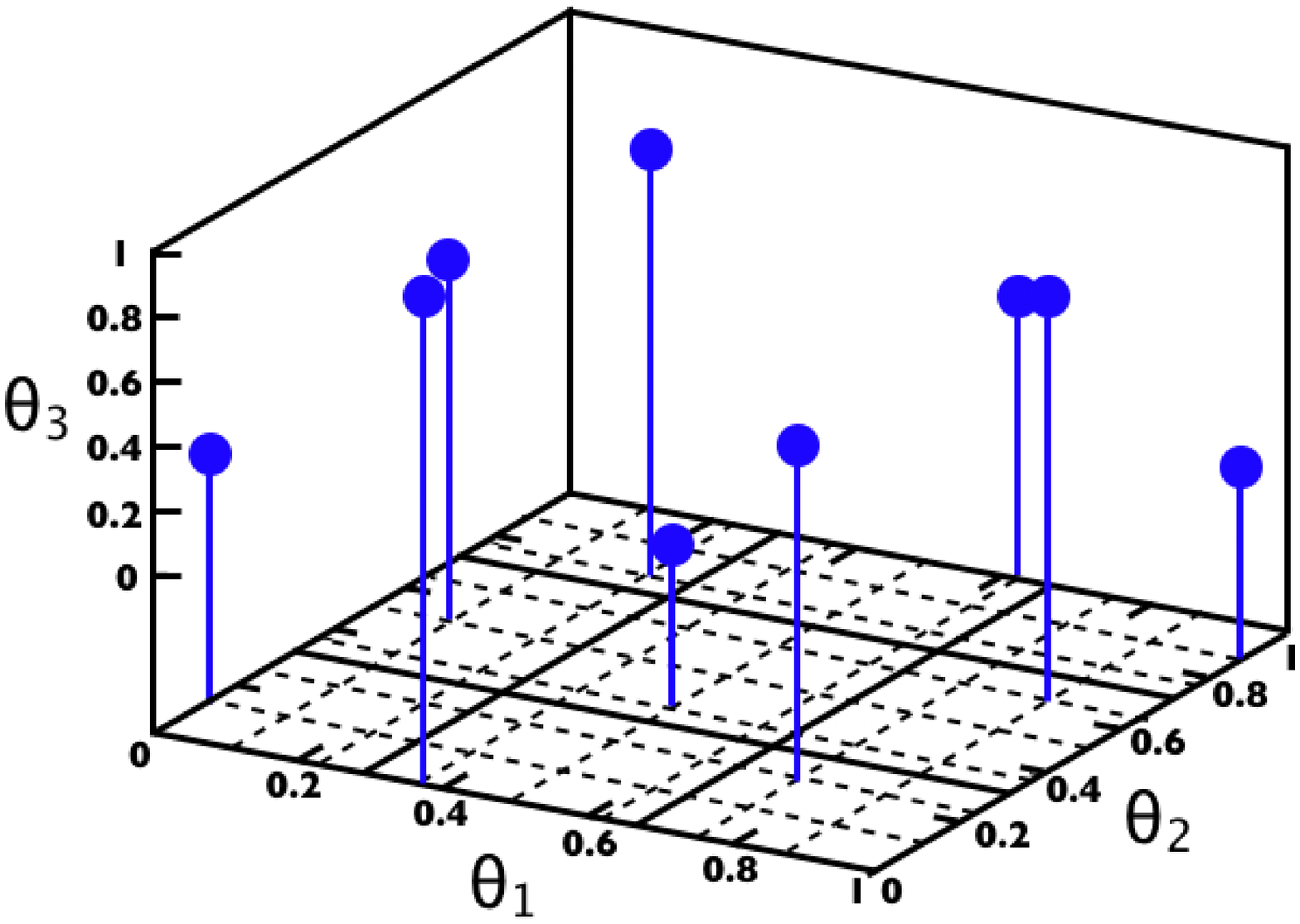}}
\caption{\label{design_3d}Left panel: an orthogonal array (OA) based
  design for 3 parameters, $\theta_1$, $\theta_2$, $\theta_3$ and nine
  sampling points. Right panel: the OA based design perturbed in such
  a way that the one-dimensional projection onto any parameter leads
  to an equally spaced distribution of sample points. The projection
  onto any two dimensions leads to a space filling design. }
\end{figure}

The above discussion points to a shortcoming of orthogonal arrays: the
number of simulation runs cannot be picked arbitrarily (e.g., choosing
$s=2$ forces a power of two for the number of runs). The other
difficulty with orthogonal arrays is that they are not easy to
construct.\footnote{A detailed description of OA designs and how to
  construct them is given in \citet{hedayat99}. A library containing a
  large number of OAs can be found at:
  http://www.research.att.com/$\sim$njas/oadir/index.html. A
  collection of C routines to generate OA designs can be found at:
  {http://lib.stat.cmu.edu/designs/oa.c}.}

Some of the preceding discussion can be best understood by studying a
specific case. Consider an example with nine sampling points ($n=9$)
and three parameters ($m=3$). If we require $r>1$ for the strength of
the design, then the relation $n=\lambda s^r$ leads automatically to
an OA design with $s=3$ levels, strength $r=2$, $\lambda=1$; an
OA$(9,3,3,2)$ design. We require that if we project our design down
onto any two-dimensional direction, the parameter space be well
covered. The left panel in Figure~\ref{design_3d} shows a possible
realization of an OA with our example specifications. The lower
triangle in Figure~\ref{design_2d} shows the three possible
two-dimensional projections of this design. This specific design is of
course not a unique solution. In matrix form it reads:

\begin{figure}[t]
\centerline{
 \includegraphics[width=3.4in]{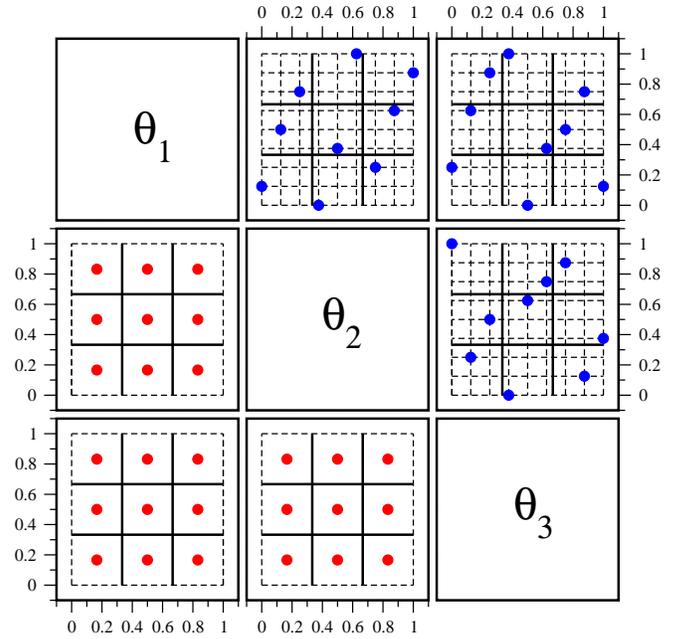}}
\caption{\label{design_2d}Projections of the design shown in
  Figure~\ref{design_3d} onto two dimensions. The lower triangle shows
  the projection of the OA design, the upper triangle of the OA-LH
  design. Note that when projected onto one dimension, the OA-LH
  design leads to an even coverage and no points lie on top of each
  other.}
\end{figure}

\begin{equation}
\label{oamatrix}
\left(\begin{array}{ccc}
\theta_1&\theta_2&\theta_3\\
\hline
1&1&1\\
2&1&3\\
3&1&2\\
1&2&2\\
2&2&1\\
3&2&3\\
1&3&3\\
2&3&2\\
3&3&1
\end{array}\right)
\stackrel{[0..1]}{\Leftrightarrow}
\left(\begin{array}{ccc}
0.166 &  0.166 &  0.166\\
0.5   &  0.166 &  0.832\\
0.832 &  0.166 &  0.5\\
0.166 &  0.5   &  0.5\\
0.5   &  0.5   &  0.166\\
0.832 &  0.5   &  0.832\\
0.166 &  0.832 &  0.832\\
0.5   &  0.832 &  0.5\\
0.832 &  0.832 &  0.166
\end{array}\right).
\end{equation}

{}From this $9\times 3$ matrix we can now verify that each of the three
$9\times 2$ sub-matrices indeed contain all possible $1\times 2$ rows
with the same frequency $\lambda=1$.  On the right hand side of
Eq.~(\ref{oamatrix}) we have simply rescaled the design points into
the normalized [0,1] space which is shown in Figures~\ref{design_3d}
and \ref{design_2d}.

\subsubsubsection{OA-LH Designs}

In order to further improve parameter space coverage, the next step --
Latin hypercube sampling -- perturbs the positions of each sampling
point from ${\bf A}$ via the following prescription: for each column
of ${\bf A}$, the $\lambda s^{r-1}$ positions with entry $k$ are
replaced by a permutation of ($k=1,\cdots,s$)

\begin{equation} 
(k-1)\lambda s^{r-1}+1,(k-1)\lambda
s^{r-1}+2,\cdots,(k-1)\lambda s^{r-1}+\lambda s^{r-1}=k\lambda s^{r-1}.
\end{equation} 

This means, in our example, that the entries for $k=1$ will be replaced
by 1,2,3, the entries for $k=2$ will be replaced by 4,5,6, and the
entries for $k=3$ by 7,8,9.  The Latin  hypercube algorithm demands 
that in every column every entry appears only once.  This ensures 
that each one dimensional projection is evenly  
covered with points and no run is replicated in the resulting  
design. The right panel in Figure~\ref{design_3d} shows a possible realization of this
in three dimensions, derived from perturbing the orthogonal array in
the left panel. The upper right triangle in Figure~\ref{design_2d}
shows the two-dimensional projection of this design. The solid lines
show the division for the orthogonal array while the dashed lines show
the additional sub-division. Note that each sample point lies on a
unique horizontal and vertical dashed line -- if the sample points are
projected down onto any one direction, the one-dimensional space would
be evenly covered. In matrix form, our OA-LH design is as follows:

\begin{equation}
\left(\begin{array}{ccc}
 1 & 2 &  3\\
 4 & 1 &  9\\
 7 & 3 &  5\\
 2 & 5 &  6\\
 5 & 4 &  1\\
 8 & 6 &  7\\
 3 & 7 &  8\\
 6 & 9 &  4\\
 9 & 8 &  2
\end{array}\right)
\stackrel{[0..1]}{\Leftrightarrow}
\left(\begin{array}{ccc}
0      & 0.125 &    0.25 \\
0.375  & 0     &    1\\
0.75   & 0.25  &    0.5\\
0.125  & 0.5   &    0.625\\
0.5    & 0.375 &    0\\
0.875  & 0.625 &    0.75\\
0.25   & 0.75  &    0.875\\
0.625  & 1     &    0.375\\
1      & 0.875 &    0.125
\end{array}\right).
\end{equation}

Note that we have replaced the entries in a random fashion and created
this design ``by hand'' -- convincing ourselves ``by eye'' that we
have good coverage in 2-D projection. \citet{leary03} suggest an
optimal strategy to ensure even better coverage of the parameter
space. These optimization strategies are often used for the projected
parameter space. In order for the points to be spread out, one has to
determine the `closeness' between them. This can be defined as the
smallest distance between any two points. A design that maximizes this
measure is said to be a maximin distance design. (For more details,
see \citealt{santner03} and Appendix~\ref{appopt}.) The designs in
\citet{HHHN} and \citet{HHHNW} combine the OA-LH based design with a
maximin distance design in each two-dimensional projection. Other
optimization methods rest on an entropy criterion based on the
minimization of -$\log|R|$, where $R$ is the covariance matrix of the
design \citep{shewry87}, or minimization of the Integrated Mean
Squared Error \citep{sachs89a}.

Our example shows just one way to realize an OA-LH design. It can be
implemented straightforwardly and leads to the desired coverage of the
parameter space. After the design has been determined in the [0,1]
parameter space, it then can be easily translated into the physical
parameter space of interest. At this point, when projected down to one
dimension, the equidistant coverage in each dimension of the parameter
space in one dimension is of course lost. However, since our
transformation is linear, we do not lose the uniformity of the
projections. Therefore it is still true that no two sample points will fall on top of
each other in projection.

\subsubsubsection{SLH designs}

As mentioned above, the major restriction of OA-LH based designs is
that they cannot be set up for an arbitrary number of simulation
runs. This is a specific point of concern, if one can only run a very
restricted number of simulations. In addition, the set-up of an OA-LH
design can be nontrivial. Very often, one has to rely on OA libraries
which are restricted in their parameters and also not always easily
available. An alternative space-filling design strategy presented by
\citet{liye00}, offers a compromise between the computing effort and
the design optimality -- the optimal symmetric Latin hypercube
designs. Following their definition, an LH design is called a
symmetric LH (SLH) design if it has the following property: For any
row $i$ of an LH design, there exists another row in the design which
is the $i$th row's reflection through the center. For example, in an
$n\times m$ Latin hypercube with levels from 1 to $n$, if
$(a_1,a_2,...,a_m)$ is one of the rows, the vector
$(n+1-a_1,n+1-a_2,...,n+1-a_m)$ should be another row in the
design. The symmetry imposes a space-filling requirement on the
designs considered up front, which carries through to all projections.
An example for an SLH design is given by:

\begin{equation}
\left(\begin{array}{ccc}
 1 & 2 &  3\\
 8 & 7 &  6\\
 4 & 1 &  2\\
 5 & 8 &  7\\
 7 & 3 &  5\\
 2 & 6 &  4\\
 3 & 4 &  1\\
 6 & 5 &  8
\end{array}\right).
\label{slh}
\end{equation}

\begin{figure}
\centerline{
 \includegraphics[width=3.3in]{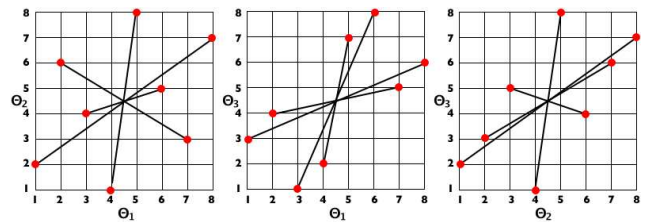}}
\caption{\label{slh_design}Two-dimensional projections of the SLH
  design given in Eq.~(\ref{slh}). The symmetric design points are
  connected to show the reflection through the center.}
\end{figure}

In this design, rows 1/2, 3/4, 5/6, and 7/8 are symmetric pairs. As
before, we do not attempt to optimize the resulting design (though the
design we use later in the paper is optimized, see below).
The two-dimensional projection of the design given in Eq.~(\ref{slh})
is shown in Figure~\ref{slh_design}.

\citet{liye00} provide an excellent discussion of optimal SLH designs,
including a description of possible algorithmic implementations and
comparison with traditional optimal LH designs. As an example, they
show that the computational effort to find an optimal LH design by
starting with an SLH design reduces roughly by a factor of ten for a
$25\times 4$ design on a workstation. As before, the SLH design is
usually optimized in the last step, often with respect to a distance
based criterion which spreads out the points in two-dimensional
projections. Two standard search algorithms for an optimal SLH design
are the columnwise-pairwise (CP) algorithm by \citet{ye98} and the
simulated annealing (SA) algorithm by \citet{mm95}. (More details
about the SA and CP algorithms are given in Appendices~\ref{SA} and
\ref{CP}, respectively). Simply put, these algorithms are based on
columnwise exchanges of entries which will keep the symmetry
properties (since the corresponding symmetric pairs are always
switched at the same time). They are iterative procedures, which will
stop after a certain pre-set optimization criterion is fulfilled or
the process is interrupted by time limitations. Very often, several
designs are produced at the same time and the most optimal kept in the
end. The SA and CP algorithms can also be used to optimize OA-LH
designs. If the OA skeleton is symmetric, then one can require the
optimal OA-LH design to be symmetric as well.

In the following, we will use an LH design optimized via a distance
criterion. The design will encompass 37 simulation runs and five
cosmological parameters.  In detail, 20 optimizations with CP and 20
with SA were carried out, and the best was chosen in the end where the
quality was measured by a distance criterion. For CP, 10 of the
designs had a symmetry requirement and the other 10 did not.  For SA,
10 of the designs had a symmetry requirement and were optimized with a
local optimization criterion, and the other 10 did not have a symmetry
requirement and were optimized with a more global optimization
criterion.  The best design from all of these came from one of the
optimizations using SA, a global optimization criterion, and no
symmetry requirement.

\subsubsection{Observational Considerations}
\label{obscon}

\begin{figure}
\centerline{
 \includegraphics[width=3.3in]{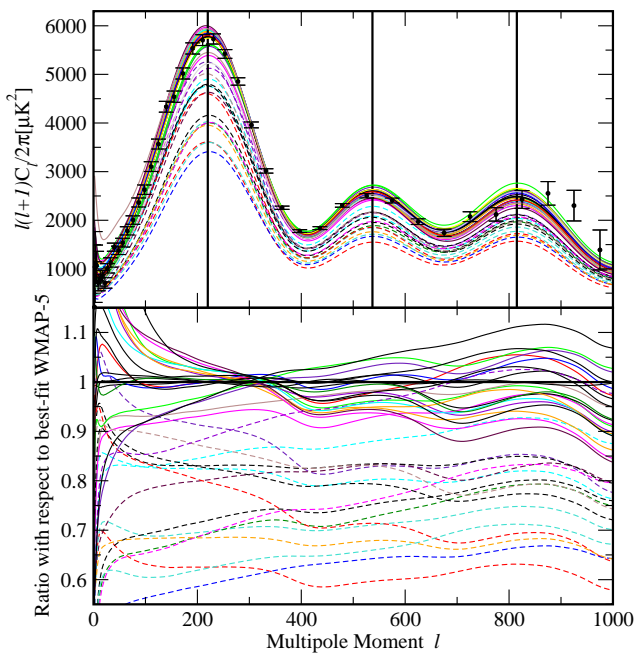}}
\caption{\label{cls}Best-fit TT power spectra for each model in
  Table~\ref{tab:basic} using the WMAP-5 results. The only parameter
  which has been optimized by minimizing $\chi^2$ is the optical depth
  $\tau$. The upper panel shows the resulting power spectra, the black
  points with error bars show WMAP-5 data points, and the thick black
  line the best-fit WMAP-5 model. The lower panel shows the residuals
  for each model with respect to the best-fit model. Some of our
  models are undernormalized, the best-fit $\tau$ being smaller than
  $0.01$ which would lead to a reionization redshift of $z<2$ and
  $\chi^2$ for these models is larger than 3000 (the $\chi^2$ for the
  best-fit WMAP-5 model is at roughly 2650). We fixed $\tau$ for those
  models at $\tau=0.01$ and show them with dashed lines.}
\end{figure}

The choice of number of active parameters depends on the available
data as well as on the chosen modeling approach. We do not insist on a
formal methodology to make this choice, but instead present practical
and conservative arguments to justify our decisions. We take as our
basic 5 parameters $\omega_m\equiv\Omega_m h^2$,
$\omega_b\equiv\Omega_b h^2$, $n_s$, $w$, and $\sigma_8$ where
$\Omega_m$ contains the contributions from the dark matter and the
baryons.  We restrict ourselves to power-law models (no running of the
spectral index), to spatially flat models without massive neutrinos
and to dark energy models with constant equation of state.  A sixth
parameter, the redshift or time, simply requires us to dump data from
each run at multiple epochs.
The choice of these parameters is dictated by the physics underlying the
matter power spectrum, and the combinations of cosmological parameters
that are well constrained by existing (primarily CMB) data
\citep[see also][]{WhiVal04}.

The effect of massive neutrinos can be roughly approximated by
decreasing $\Omega_m$ \citep{BHHT}.  We do not expect any
realistic dark energy model to have a constant equation of state, but
we wanted to begin with the most restrictive parameter space in order
to validate our methods.  The next generation of experiments will pose
at best weak constraints on any time variation of $w$, and in this
sense our constant $w$ can be thought of as an appropriate average of
$w(z)$.  Using growth matching techniques
\citep{WhiVal04,LinWhi05,FraLewLin07} one can map the power spectrum
from a complex $w(z)$ onto one with a constant $w$ with reasonable
accuracy.

The normalization is another area where choices need to be made.
Historically the amplitude of the power spectrum was set by
$\sigma_8$, the amplitude of the {\it linear theory\/} matter power
spectrum smoothed with a top-hat on scales of $8\,h^{-1}$Mpc
\begin{equation}\label{linp}
  \sigma_8^2 \equiv \int \frac{dk}{k}\ \Delta^2_{\rm lin}(k)
  \left. \left[ \frac{3j_1(kR)}{kR} \right]^2 \right|_{R=8\,h^{-1}{\rm Mpc}}
  \quad ,
\end{equation}
with the linear power spectrum being defined as
\begin{equation}
\label{plin}
\Delta_{\rm lin}^2(k)\equiv \frac{k^3P_{\rm lin}(k)}{2\pi^2}.
\end{equation}
This scale and normalization were chosen because the fluctuations of
counts of $L_\star$ galaxies in cells of this size is close to unity.
With the advent of the {\sl COBE\/} data it became common to quote the
normalization at horizon scales, e.g.~\citet{BunWhi97}.  As CMB data
improved, the pivot point was shifted to smaller scales, closer to the
middle of the range over which the spectrum is probed and where the
normalization is best determined.  In order to make closer connection
with the initial fluctuations, the amplitude not of the matter power
spectrum but of the curvature or potential fluctuations has been adopted.
These differ mostly by factors of growth and $\Omega_m$.  Anticipating
future advances, $k_p=0.002\,{\rm Mpc}^{-1}$ was selected for the most
recent CMB analysis by \citet{WMAP5} and the rms curvature fluctuation
on this scale is now most commonly used as a normalization.  With
present CMB data the biggest uncertainties in the normalization are
the near degeneracy with the optical depth, $\tau$, and the uncertain
growth of perturbations at low redshift due to the unknown equation of
state of the dark energy, e.g.~\citet{Whi06}.

\begin{table*}[t]
\begin{center} 
\caption{\label{tab:basic} Parameters for 38 Models}
\begin{tabular}{cccccc|cccccc}
\tableline\tableline
\#& $\omega_m$ & $\omega_b$ & $n_s$ & $-w$ & $\sigma_8$ &
\#& $\omega_m$ & $\omega_b$ & $n_s$ & $-w$ & $\sigma_8$ \\
\hline 
0 & 0.1296 &  0.0224 & 0.9700 & 1.000 &  0.8000 &
19 & 0.1279 & 0.0232 & 0.8629 & 1.184 & 0.6159 \\ 
1 & 0.1539 & 0.0231 & 0.9468 & 0.816 & 0.8161 &
20 & 0.1290 & 0.0220 & 1.0242 & 0.797 & 0.7972 \\ 
2 & 0.1460 & 0.0227 & 0.8952 & 0.758 & 0.8548 &
21 & 0.1335 & 0.0221 & 1.0371 & 1.165 & 0.6563 \\ 
3 & 0.1324 & 0.0235 & 0.9984 & 0.874 & 0.8484 & 
22 & 0.1505 & 0.0225 & 1.0500 & 1.107 & 0.7678 \\ 
4 & 0.1381 & 0.0227 & 0.9339 & 1.087 & 0.7000 & 
23 & 0.1211 & 0.0220 & 0.9016 & 1.261 & 0.6664 \\ 
5 & 0.1358 & 0.0216 & 0.9726 & 1.242 & 0.8226 & 
24 & 0.1302 & 0.0226 & 0.9532 & 1.300 & 0.6644 \\ 
6 & 0.1516 & 0.0229 & 0.9145 & 1.223 & 0.6705 & 
25 & 0.1494 & 0.0217 & 1.0113 & 0.719 & 0.7398 \\ 
7 & 0.1268 & 0.0223 & 0.9210 & 0.700 & 0.7474 & 
26 & 0.1347 & 0.0232 & 0.9081 & 0.952 & 0.7995 \\ 
8 & 0.1448 & 0.0223 & 0.9855 & 1.203 & 0.8090 & 
27 & 0.1369 & 0.0224 & 0.8500 & 0.836 & 0.7111 \\ 
9 & 0.1392 & 0.0234 & 0.9790 & 0.739 & 0.6692 & 
28 & 0.1527 & 0.0222 & 0.8694 & 0.932 & 0.8068\\ 
10 & 0.1403 & 0.0218 & 0.8565 & 0.990 & 0.7556 & 
29 & 0.1256 & 0.0228 & 1.0435 & 0.913 & 0.7087 \\ 
11 & 0.1437 & 0.0234 & 0.8823 & 1.126 & 0.7276 & 
30 & 0.1234 & 0.0230 & 0.8758 & 0.777 & 0.6739 \\ 
12 & 0.1223 & 0.0225 & 1.0048 & 0.971 & 0.6271 & 
31 & 0.1550 & 0.0219 & 0.9919 & 1.068 & 0.7041 \\ 
13 & 0.1482 & 0.0221 & 0.9597 & 0.855 & 0.6508 & 
32 & 0.1200 & 0.0229 & 0.9661 & 1.048 & 0.7556\\
14 & 0.1471 & 0.0233 & 1.0306 & 1.010 & 0.7075 & 
33 & 0.1399 & 0.0225 & 1.0407 & 1.147 & 0.8645 \\ 
15 & 0.1415 & 0.0230 & 1.0177 & 1.281 & 0.7692 & 
34 & 0.1497 & 0.0227 & 0.9239 & 1.000 & 0.8734 \\  
16 & 0.1245 & 0.0218 & 0.9403 & 1.145 & 0.7437 & 
35 & 0.1485 & 0.0221 & 0.9604 & 0.853 & 0.8822 \\ 
17 & 0.1426 & 0.0215 & 0.9274 & 0.893 & 0.6865 & 
36 & 0.1216 & 0.0233 & 0.9387 & 0.706 & 0.8911 \\
18 & 0.1313 & 0.0216 & 0.8887 & 1.029 & 0.6440 &
37 & 0.1495 & 0.0228 & 1.0233 & 1.294 & 0.9000 \\
\tableline\tableline

\vspace{-1.5cm}

\tablecomments{The five basic parameters for the 37 models that
  specify the simulation design  and model 0 which we use as an independent  
check. See text for definitions.}
\end{tabular}
\end{center}
\end{table*}

\begin{table*}
\begin{center} 
\caption{\label{tab:derived} Derived Parameters for 38 Models}
\begin{tabular}{cccccc|cccccc}
\tableline\tableline
\#& $\Omega_m$ & $\Omega_{\rm de}$ & $h$ & $d_{\rm ls}$ &
$\tau(\chi^2)$ &
\#& $\Omega_m$ & $\Omega_{\rm de}$ & $h$ & $d_{\rm ls}$ & $\tau(\chi^2)$ \\
 
0 & 0.2500 & 0.7500 & 0.7200 & 14.24 & 0.1 &
19 & 0.1940 & 0.8060 & 0.8120 & 14.24 & < 0.01 (7712)\\
1 & 0.4307 & 0.5693 & 0.5977 & 13.59 & 0.064 &
20 & 0.3109 & 0.6891 & 0.6442 & 14.27 & 0.15  \\
 2 & 0.4095 & 0.5905 & 0.5970 & 13.80 & 0.205 &
21 & 0.2312 & 0.7688 & 0.7601 & 14.14 & < 0.01 (21579)\\
 3 & 0.2895 & 0.7105 & 0.6763 & 14.10 & 0.19 &
22 & 0.3317 & 0.6683 & 0.6736 & 13.70 & < 0.01 (11139) \\
 4 & 0.2660 & 0.7340 & 0.7204 & 13.99 & < 0.01 (5569) &
23 & 0.1602 & 0.8398 & 0.8694 & 14.48 & < 0.01 (4478)\\
 5 & 0.2309 & 0.7691 & 0.7669 & 14.11 & < 0.01 (2756) &
24 & 0.1854 & 0.8146 & 0.8380 & 14.21 & < 0.01 (13138) \\
 6 & 0.3059 & 0.6941 & 0.7040 & 13.66 & < 0.01 (19318) &
25 & 0.4558 & 0.5442 & 0.5724 & 13.76 & < 0.01 (3033) \\
 7 & 0.3310 & 0.6690 & 0.6189 & 14.31 & 0.225 &
26 & 0.2804 & 0.7196 & 0.6931 & 14.05 & 0.14 \\
 8 & 0.2780 & 0.7220 & 0.7218 & 13.84 & < 0.01 (4320) &
27 & 0.3357 & 0.6643 & 0.6387 & 14.04 & 0.08 \\
 9 & 0.3707 & 0.6293 & 0.6127 & 13.93 & < 0.015 (2845)&
28 & 0.3988 & 0.6012 & 0.6189 & 13.66 & 0.06 \\
10 & 0.3131 & 0.6869 & 0.6695 & 13.98 & 0.05 &
29 & 0.2516 & 0.7484 & 0.7067 & 14.32 & < 0.01 (2809)\\
11 & 0.2790 & 0.7210 & 0.7177 & 13.82 & < 0.01 (3928) &
30 & 0.2810 & 0.7190 & 0.6626 & 14.37 & 0.155 \\
12 & 0.2235 & 0.7765 & 0.7396 & 14.43 & < 0.01 (5901) &
31 & 0.3791 & 0.6209 & 0.6394 & 13.62 & < 0.01 (17774) \\
13 & 0.3974 & 0.6026 & 0.6107 & 13.77 & < 0.01 (11549) &
32 & 0.1922 & 0.8078 & 0.7901 & 14.47 & 0.115 \\
14 & 0.3289 & 0.6711 & 0.6688 & 13.74 & < 0.01 (11803) &
33 & 0.2634 & 0.7366 & 0.7286 & 13.96 & <0.01 (2829)\\
15 & 0.2363 & 0.7637 & 0.7737 & 13.89 & < 0.01 (9905) &
34 & 0.3532 & 0.6468 & 0.6510 & 13.71 & 0.1 \\
16 & 0.1981 & 0.8019 & 0.7929 & 14.40 & 0.025 &
35 & 0.3990 & 0.6010 & 0.6100 & 13.77 & 0.135 \\
17 & 0.3586 & 0.6414 & 0.6305 & 13.94 & < 0.01 (5012) &
36 & 0.2949 & 0.7051 & 0.6421 & 14.41 & 0.455 \\
18 & 0.2578 & 0.7422 & 0.7136 & 14.22 & < 0.01 (5641) &
37 & 0.2796 & 0.7204 & 0.7313 & 13.71 & < 0.01 (2971)\\
\tableline
\tableline

\vspace{-1.5cm}

\tablecomments{The derived parameters, obtained from the basic parameters
  listed in Table \protect\ref{tab:basic} by applying the constraint on
  $\ell_A$.  Only for model 0 is the Hubble parameter picked by hand. The
  distance to last scattering is in Gpc, all other parameters are
  dimensionless.  See text for details.}
\end{tabular}
\end{center}
\end{table*}

For our purposes, however, a normalization tied to the present day matter
power spectrum and close to the nonlinear scale has many advantages.
Rather than introduce yet another convention, we therefore choose to use
$\sigma_8$ as our normalization parameter.  Of course, since all of the
parameters of the models are specified one can compute any other parameter
for our models.  As an example, we have evaluated for each of the models
given in Table \ref{tab:basic} the best-fit value for $\tau$ using the
likelihood code provided by the WMAP-5 team.  The resulting TT power spectra
are shown in Figure \ref{cls} as well as their ratios with respect to the
best-fit WMAP-5 model.  Some of our models are undernormalized and the
resulting $\tau$ is smaller than 0.01 leading to reionization redshifts of
$z<2$.  This undernormalization however is not a concern: we chose the
models to cover the parameter space well overall and not to provide fits
close to the concordance cosmology.  We provide the best-fit values for
$\tau$ in Table \ref{tab:derived}.

\begin{figure}[b]
\centerline{
 \includegraphics[width=2.5in]{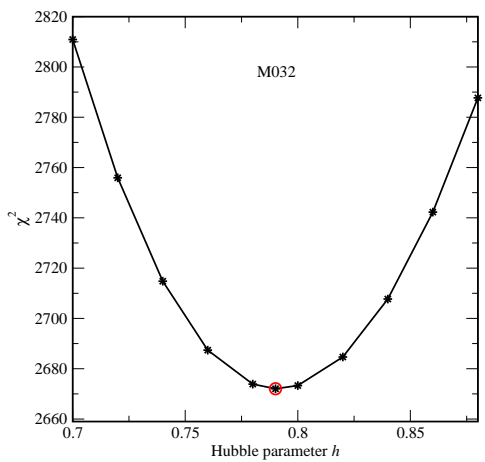}}
\caption{\label{fig:hubble} Sweep through $h$ for model 32. The red
  circle marks the estimate for the Hubble parameter from assuming
  perfect knowledge of $\ell_A$, in excellent agreement with the
  result from the WMAP-5 likelihood for the best-fit value of $h$ for
  this model.}
\end{figure}

{}From the WMAP 5-year data, in combination with BAO, we have\footnote{See
http://lambda.gsfc.nasa.gov/}
\begin{equation}
\begin{array}{c}
  \omega_m = 0.1347\pm 0.0040\quad (3\%), \\
  \omega_b = 0.0227\pm 0.0006\quad (3\%), \\
  n_s      = 0.9610\pm 0.0140\quad (2\%).
\end{array}
\end{equation}
Current data restrict a constant equation of state for the dark energy
to $w=-1$ with roughly 10\% accuracy (for very recent results from
supernovae see, e.g., \citealt{kow08}; for weak lensing see, e.g.,
\citealt{kilb08}; and for the latest constraints from clusters of
galaxies, see \citealt{vik08}). Recent determinations put the
normalization in the range $0.7-0.9$ with still rather large
uncertainties (see, e.g., \citealt{vik08} for constraints from
clusters, \citealt{voevod04} for a low estimate from clusters,
\citealt{teg07} for constraints from combined CMB and large scale
structure data, and \citealt{evrad08} for an extended discussion of
recent results). Considering all these constraints and their
uncertainties, we choose our sample space boundaries to be
\begin{equation}
\label{priorrange}
\begin{array}{c}
  0.120 < \omega_m < 0.155, \\
  0.0215 < \omega_b < 0.0235, \\
  0.85  < n_s      < 1.05, \\
  -1.30 < w        <-0.70, \\
  0.61  < \sigma_8 < 0.9,
\end{array}
\end{equation}
as shown in Table \ref{tab:basic}.

In order to solve for the full set of cosmological parameters we
impose the CMB constraint that $\ell_A\equiv \pi d_{\rm
  ls}/r_s=302.4$, where $d_{\rm ls}$ is the distance to the last
scattering surface and $r_s$ is the sound horizon.  Observationally
this is known to $0.3\%$, and models which significantly violate this
equality are poor fits to the CMB data (see Figure \ref{fig:hubble}).
Unfortunately the sound horizon, like the epoch of last scattering,
can be defined in a number of different ways which differ subtly.
Specifically we use the fit to the redshift of last scattering of
Eq.~(23) of \citet{Damp} and use Eq.~(B6) of \citet{HuSug95} for the
sound horizon.  With these choices we find models with $\omega_m$ and
$\omega_b$ in the range preferred by WMAP do indeed provide good fits
to the WMAP data. This is demonstrated for model 32 in Figure
\ref{fig:hubble}.

The procedure is then as follows.  For every specified $\omega_m$ and
$\omega_b$ we compute $r_s$ and hence the required $d_{\rm ls}$ to fit
the $\ell_A$ constraint.  We adjust $h$, at fixed spatial curvature,
$w$, and $\omega_m$, until the model reproduces the required $d_{\rm
  ls}$.  Knowing $h$ and $\omega_m$ then gives us $\Omega_m$ and hence
$\Omega_{\rm de}$, as shown in Table \ref{tab:derived}.

Finally, we generated a model `0' which has parameters close to the
current best fit from CMB and large-scale structure
(e.g.,~\citealt{WMAP5}).  This model has $\Omega_m=0.25$, $\Omega_{\rm
  de}=0.75$, $\omega_b=0.0224$, $n_s=0.97$, $h=0.72$, and
$\sigma_8=0.8$ and can be used as an independent check of the
interpolation in the range of most interest.

\subsubsection{The Resulting Design}

\begin{figure}[b]
\centerline{
 \includegraphics[width=3.2in]{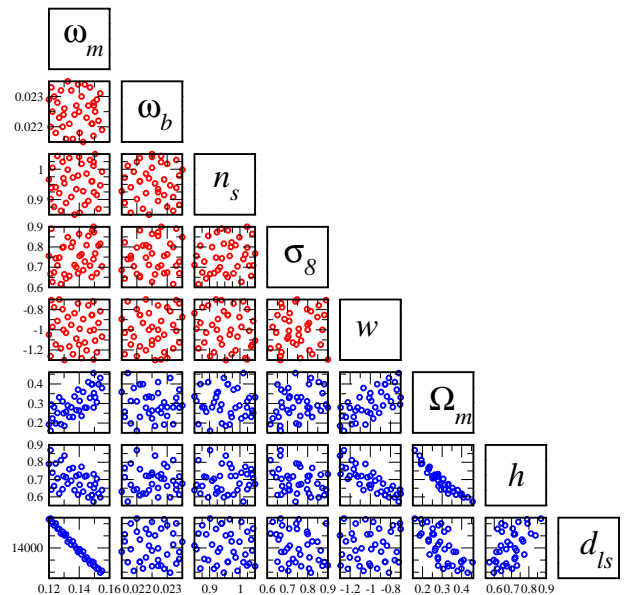}}
\caption{\label{design} Final simulation design: The five parameters
  under consideration are shown in red, projected onto two
  dimensions. The blue points show three derived parameters:
  $\Omega_m$, $h$, and $d_{\rm ls}$.}
\end{figure}

Based on the above considerations, we can now generate a space-filling
design for the five parameters of interest. We restrict ourselves to
37 cosmologies and will show in the remainder of the paper that this
number is indeed sufficient to generate an accurate emulator. For
model 0 we pick a standard LCDM model for which we chose the Hubble
parameter by hand (although $h=0.72$ is very close to the result we
would obtain if we would derive it from $d_{\rm ls}$).  The resulting
cosmological models are listed in Table~\ref{tab:basic} where we give
the values for the basic parameters. In Table~\ref{tab:derived} we
give in addition a few derived parameters: $\Omega_m$, $\Omega_{\rm
  de}$ (recall that flatness is assumed), $h$ as derived from our
constraint on $\ell_A$, $d_{\rm ls}$, and $\tau$.

\begin{figure}[t]
\centerline{
 \includegraphics[width=3.3in]{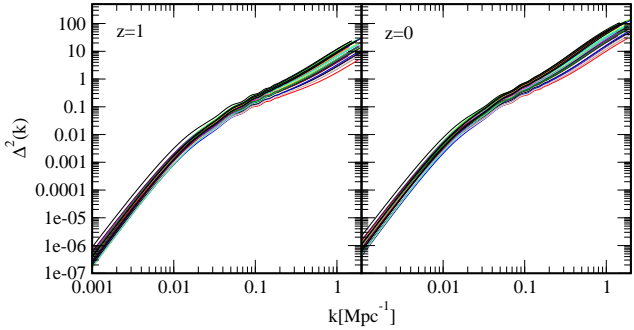}}
\caption{\label{deltasq}Dimensionless power spectra as given by {\sc
    HaloFit} for the 38 cosmologies specified in Table~\ref{tab:basic}
  at $z=1$ (left panel) and $z=0$ (right panel).}
\end{figure}

\begin{figure*}
\centerline{
 \includegraphics[width=5.5in]{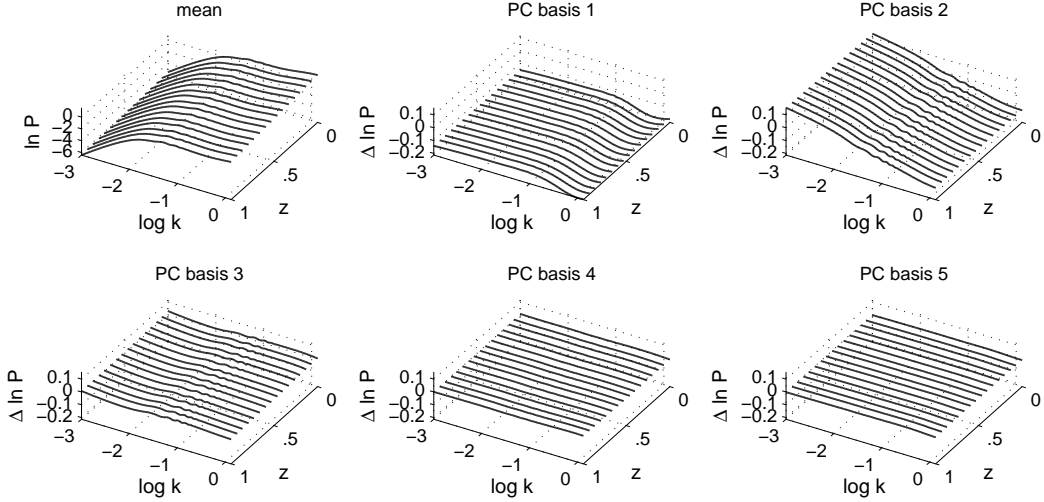}}
\caption{\label{basis}Mean (upper left corner) and five principal
  component (PC) bases derived from the output from the 38 {\sc
    HaloFit} power spectra. The third axis shows the time evolution of
  the mean and the five principal components between redshift $z=1$
  and $z=0$. While the first two principal component bases show
  significant variation, the fourth and fifth are already almost flat,
  indicating that the inclusion of higher principal components would
  not improve the quality of the emulator and the underlying GP
  model.}
\end{figure*}

The two-dimensional projections of the design are shown in Figure
\ref{design}.  The upper part of the triangle shows the five input
parameters in red, demonstrating a good sampling of the parameter
space.  The blue points show projections of three of the derived
parameters, $\Omega_m$, $h$, and $d_{\rm ls}$.

The key statistical observable discussed in this paper is the density
fluctuation power spectrum $P(k)$, the (ensemble-averaged) Fourier
transform of the two-point density correlation function.  In
dimensionless form, the power spectrum can be written as
\begin{equation}\label{nonlinp}
  \Delta^2(k)\equiv \frac{k^3P(k)}{2\pi^2},
\end{equation}
equivalent to the linear power spectrum in Eq.~(\ref{linp}).
Figure~\ref{deltasq} shows the resulting dimensionless {\sc HaloFit\/}
power spectra for the 38 cosmological models at $z=1$ (left panel) and
at $z=0$ (right panel)\footnote{Note that we are using Mpc$^{-1}$
  units for $k$ in this paper, not $h\,{\rm Mpc}^{-1}$ as in
  \citet{Heitmann08}.  We omit the $h$ since the underlying physics
  determines the shape of the power spectrum better in Mpc$^{-1}$
  units than $h\,{\rm Mpc}^{-1}$ units \citep[e.g.][]{Whi06}.}.
Overall, the parameter space is well covered by these 38 models, the
coverage being sufficient to accommodate upcoming weak lensing survey
measurements.

\subsection{Emulation}
\label{gpm}

After specifying the design, the next task is to construct the
emulator for predicting the nonlinear matter power spectrum within the
parameter priors specified in the design.  For an in-depth
mathematical description of building such an emulator in the
cosmological context we refer the reader to \citet{HHHNW} and
\citet{SKHHHN}.  Here we explain the major ideas behind the process
and show explicitly the emulator performance for our 37 model design.
As discussed in the Introduction, we use {\sc HaloFit\/} as a proxy
for the full numerical simulations as a convenient foil to demonstrate
and to test the overall procedure.

Before constructing the emulator, it is useful to find the
best-possible representation of the power spectrum. The major aim is
to find a representation that preserves the smoothness of the power
spectrum but at the same time enhances important features, such as the
baryon acoustic oscillations. It turns out to be more convenient to
model the power spectrum as
\begin{equation}
  {\cal P}(k,z;\theta)= \ln\left\{ \frac{\Delta^2(k,z)}{2\pi k^{3/2}}
  \right\}  
\end{equation}
than to work with $\Delta^2(k,z;\theta)$ directly (the connection
between $\Delta^2(k)$ and $P(k)$ is given in Eq.~(\ref{nonlinp}),
$\theta$ denotes the cosmological parameters varied).  This
transformation reduces the total dynamic range as well as enhances the
baryon acoustic oscillation features in the power spectrum.

In order to construct the emulator, we represent the scaled power
spectrum ${\cal P}(k,z;\theta)$ using an $n_{\cal P}$-dimensional
basis: 
\begin{equation}
  {\cal P}(k,z;\theta)=\mu_{\cal P}(k,z) + \sum_{i=1}^{n_{\cal P}}
    \phi_i(k,z)w_i(\theta)+\epsilon,
    \qquad\theta\in [0,1]^{n_\theta},
\label{model}
\end{equation}
where the $\phi_i$ are the basis functions and the $w_i$ are the
corresponding weights.  We have stored each power spectrum at 200 $k$
values between $-3\leq \log_{10} k \leq 0.12$ and 100 $z$ values
between $0 \leq z \leq 1$ Therefore, ${\cal P}(k,z;\theta)$ represents
the power spectra, over a $200 \times 100$ grid of $k$ and $z$ values.
Over this grid, the values ${\cal P}(k,z;\theta)$ are determined by
the five cosmological parameters denoted by $\theta$.  The
dimensionality $n_{\cal P}$ refers to the number of orthogonal basis
vectors $\{\phi_1(k,z),...,\phi_{n_{\cal P}}(k,z)\}$. As we will show
later, $n_{\cal P}=5$ turns out to be an adequate choice for the
present application. The parameter $n_\theta$ is the dimensionality of
our parameter space -- with 5 cosmological parameters we have
$n_\theta=5$ (that $n_{\cal P}=n_\theta$ here is a coincidence).
As mentioned earlier, it is convenient to map the parameter ranges into
$[0,1]$ via a linear transformation.
The last term in Eq.~(\ref{model}), $\epsilon$, is the error term.
Our main tasks in building the emulator are now:
\begin{itemize}
\item Find a suitable set of orthogonal basis vectors
  $\phi_i(k,z)$. In our case, a principal components basis turns out
  to be an efficient representation, but alternative representations
  may be employed.  We found it convenient to construct the basis vectors
  in $(k,z)$ space, though one could also build a separate basis for each
  redshift.
\item Model the weights $w_i(\theta)$ as smooth functions of the
  underlying parameters, $\theta$. Our choice of GP models is based on
  their success in representing functions that change smoothly
  with parameter variation, e.g., the variation of the power spectrum
  as a function of cosmological parameters.
\end{itemize}

\subsubsection{The Principal Component Basis}
\label{sec:pca}

Before we determine the basis vectors to model the simulation output
for the power spectrum ${\cal P}(k,z;\theta)$, we standardize the
simulation power spectra in the following way. We first center the
power spectra around their mean, given by $\mu_{\cal P}(k,z) =
\frac{1}{37}\sum_{j=1}^{37}{\cal P}_j$. The resulting mean as a
function of redshift $z$ and wavenumber $k$ is shown in the upper left
corner of Figure~\ref{basis}. (Remember that we divide $\Delta^2(k,z)$
by $2\pi k^{3/2}$ leading to the flattening of the power spectrum at
high $k$.)  After having centered the simulations around the mean, we
scale the output by a single value for each $k$ and $z$ so that its
variance is one.

The next step is the principal component analysis (PCA) to determine
the orthogonal basis vectors $\phi_i(k,z)$ for modeling the simulation
output for the power spectra following Eq.~(\ref{model}). To carry out
this step, we write the standardized power spectra as an $n_{kz}\times
m$ matrix, where $n_{kz}=20000$ is the number of support points for
each power spectrum over the $200 \times 100$ $k$-$z$ grid, and $m=37$
is the number of cosmological models\footnote{As mentioned above, we
could also determine the orthogonal basis vectors separately for each
$z$ output. In this case, we would obtain 100 matrices of size
$200\times 37$ each.  One could fit separate GP's for each $z$, but
such a model will require 500 GPs, instead of 5.  While we have not
tried fitting separate GPs at each z, we expect the resulting emulator
performance will be similar.  Using basis elements over the $(k,z)$
support results in a simpler model and much easier computations for
parameter estimation and emulation.}.  The $n_{kz}\times m$ matrix
reads:
\begin{equation}
  Y_{\rm sims}=[{\cal P}_1;...;{\cal P}_{37}].
\end{equation}
Following \citet{HHHNW}, we apply a singular value decomposition to the
simulation output matrix $Y_{\rm sims}$ giving 
\begin{equation}
  Y_{\rm sims}=UDV^T,
\end{equation}
where U is an $n_{kz}\times m$ orthogonal matrix, D is a diagonal $m\times m$
matrix holding the singular values, and $V$ is an $m\times m$ orthonormal
matrix.  The PC basis matrix $\Phi_{\cal P}$ is now defined to be the first
$n_{\cal P}$ columns of $[UD/\sqrt{m}]$ and the principal component weights
are given by $W=[\sqrt{m}V]$.  Here the $i^{\rm th}$ column of $W$ (given by 
$w^*_i = (w^*_{i1},\ldots,w^*_{im})'$) holds the weights corresponding to 
the basis function $\phi_i(k,z)$ for the $m=37$ simulations run at 
cosmologies $\theta^*_1,\ldots,\theta^*_m$.  The star indicates quantities
derived from the $m$ simulations.

In order to model the nonlinear matter power spectrum, we find that five
principal components are sufficient to capture all information.
Therefore we have $n_{\cal P}=5$ and
$\Phi_{\cal P}=[\phi_1;\phi_2;\phi_3;\phi_4;\phi_5]$.
The resulting five PC bases are shown in Figure \ref{basis} as a function
of $z$ and $k$.  The fourth and the fifth components are already very flat
-- increasing the dimensionality further would not improve the quality of
the data representation.

\subsubsection{Gaussian Process Modeling}
\label{sec:gpm}

The final step is to model the PC weight functions $w_i(\theta),\,
i=1,\ldots,n_{\cal P},$ in Eq.~(\ref{model}) conditioned on the known
results from the 37 cosmological models. We will employ Gaussian
process modeling for this task. Gaussian process modeling is
a nonparametric regression approach particularly well suited for
interpolation of smooth functions over a parameter space. As mentioned
previously, GP models are local interpolators and work well with
space-filling sampling techniques. The GP (also called Gaussian random
functions) is simply a generalization of the Gaussian probability
distribution, extending the notion of a Gaussian distribution over
scalar or vector random variables to function spaces. (For an
excellent introduction to Gaussian processes, see
\citealt{Rasmussen06}.) The Gaussian distribution is specified by a
scalar mean $\mu$ or a mean vector and a covariance matrix --
extending this to the GP leads to a specification of the GP by a mean
function and a covariance function.  

In order to build an emulator from a GP three steps have to be performed:

\begin{itemize}
\item {Specification of the Gaussian Process:} Define a GP model which
  is determined by its mean and covariance specification. A priori,
  realizations of functions produced from this model should cover a
  (specified) range of possibilities for the simulation response, here
  the PC weights $w_i(\theta)$.

\item {Statistical Parameter Estimation:} The statistical parameters
  controlling the GP model are estimated from the known simulation
  outputs via either maximum likelihood or Bayesian methods, in order
  to obtain good predictive performance from the emulator.

\item {The Conditional Process:} Once the statistical parameters have
  been estimated, the GP model is fully specified. Because of its
  Gaussian form, it can now be conditioned analytically to the
  simulation output: The conditional process produces random functions
  which are constrained to pass through the observed output and give
  predictions, with uncertainties, at untried $\theta$ values. 
\end{itemize}

We now describe each of these steps for a single one of the
$w_i(\theta)$'s; for simplicity we will ignore the subscript. From the
SVD decomposition of the simulation output described in Section~\ref{sec:pca}, we also
have the observed weights obtained from the 37 simulations which are
in the 37-vector $w^*$. The complete procedure to obtain the emulator
for the power spectra is more involved, in part due to the higher
dimensionality of the problem. The interested reader can find details
on this procedure in Appendix~\ref{app_gp}.

\subsubsubsection{Specification of the Gaussian Process}

The first step is to define the GP from which we can generate random
function realizations $w({\theta})$. For a GP, any finite restriction
of $w(\theta)$ has a multivariate Gaussian distribution. A priori, at
any single $\theta$, we take $w(\theta)$ to be mean-zero (although
this is not necessary), with variance $\sigma^2=\lambda^{-1}_w$. At
two points $\theta$ and ${\theta'}$ the $w(\theta)$ and $w(\theta')$
are correlated:
\begin{equation}
  {\rm Corr}\left[ w(\theta),w({\theta'})\right] \equiv R(\theta,\theta').
\end{equation}
The correlation function $R$ as defined above is critical to the GP
approach. In  addition to being positive definite, it typically has
the following attributes: (i) for  $\theta={\theta'}$ it is unity, so
that replicates are perfectly correlated; (ii) it is large when
$\theta\simeq {\theta'}$, capturing the notion that two nearby points
are highly correlated, (iii) it  is small when $\theta$ is far away
from ${\theta'}$ so that far separated points are essentially
uncorrelated. A commonly used form for the correlation function is
given by 
\begin{equation}\label{sigma}
  R({\theta},{\theta'};\vartheta)=\prod^{n_{\theta}}_{j=1}
  \exp\left(-\vartheta_j|\theta_j-\theta_j'|^{p_j}\right).
\end{equation}
The corresponding GP covariance function is given by
$\Sigma(\theta,\theta')=\sigma^2\,R({\theta},{\theta'};\vartheta)$.
As previously, the dimension of the problem is denoted by $n_\theta$.
Large values of $\vartheta_j$ (analog to an inverse correlation
length) allow for more complexity in the $j^{\rm th}$ component
direction for the function we want to model. The value for $p$ is
typically set to be either one or two. The choice $p=2$ yields very
smooth realizations with infinitely many derivatives, while $p=1$
leads to rougher realizations suited to modeling continuous but
non-differentiable (stochastic) functions. Throughout this paper we
will use $p=2$.    

\subsubsubsection{Statistical Parameter Estimation}

We treat the simulation output $w^*$, produced at the 37 input settings,
as a partial realization of the prior GP model.  Therefore, given the
covariance parameters $\sigma^2$ and $\vartheta$, $w^*$ is a draw
from a $m=37$-dimensional Gaussian distribution
\begin{eqnarray}
\label{like}
  &&p(w^*|\sigma^2,\vartheta_1,\dots,\vartheta_{n_\theta})=\nonumber\\
  &&(2\pi\sigma^2)^{-m/2} \det(R)^{-1/2}
  \exp\left\{ -\frac{1}{2\sigma^2} {w^*}' R(\theta^*;\vartheta)^{-1}
    w^* \right\}. 
\end{eqnarray}
Here $R(\theta^*;\vartheta)$ denotes the $37 \times 37$ correlation
matrix obtained by applying Eqn.~(\ref{sigma}) to each pair of simulated
cosmologies in $\theta^*$. 

Maximum likelihood estimates for $\sigma^2$ and $\vartheta$ can be
obtained by finding the parameter values
$(\hat{\sigma}^2,\hat{\vartheta})$ that maximize Eqn.~(\ref{like})
above.  Conditional on knowing the correlation parameters $\vartheta$,
the likelihood is maximized for $\sigma^2$ at
\begin{equation}
\hat{\sigma}^2 =
\frac{1}{m} {w^*}' R(\theta^*;\vartheta)^{-1} w^*.  
\end{equation}
Maximizing with respect to the components of $\vartheta$ must be done
numerically. Once estimated, we can insert these values in
Eqn.~(\ref{like}) and treat the GP specification as fully known.

An alternative is to use a Bayesian approach as detailed in
\citet{HHHNW} or, more generally in \citet{hgwr08}, to find the
optimal parameters of the GP.  While such an approach requires
additional prior specifications and computational effort, it does
account for uncertainties in the covariance parameters and the
resulting prediction.  This can be important if the emulator is used
along with physical observations to estimate cosmological parameters.
The resulting emulator predictions are not very sensitive to either
approach.  The analysis presented in this paper follows the Bayesian
approach detailed in the references above.  Therefore, we obtain a
posterior distribution for the covariance parameters, rather than a
point estimate as in maximum likelihood. Figure~\ref{box} shows
boxplots of the marginal posterior distribution for each of the
correlation parameters.  We provide details on this approach in
Appendix~\ref{app_gp}.

\subsubsubsection{The Conditional Process}

We can now fix $\sigma^2$ and $\vartheta$ at their estimated values,
and treat them as known. The next step is to construct the conditional
process for $w(\theta)$ given the output $w^*$ of the 37
simulations. This new, conditional process is also a GP, but
realizations of it are random functions that are constrained to
interpolate the simulation output. For any input setting $\theta$, the
conditional process describes $w(\theta)$ with a normal distribution:
\begin{eqnarray}
  &&p(w(\theta)|w^*,\hat{\sigma}^2,\hat{\vartheta})=
  (2\pi\sigma_{\theta}^2)^{-m/2} \det(R)^{-1/2}\nonumber\\  
&&\times \exp\left\{ -\frac{1}{2\sigma_{\theta}^2}
  (w(\theta)-\mu_{\theta})^TR(w(\theta)-\mu_{\theta})\right\}.    
\end{eqnarray}
Here $\mu_\theta$ and $\sigma^2_\theta$ are determined by the
parameters $\sigma^2$, $\vartheta$, by the simulation output 
$w^*$, and by the componentwise distances between the new
prediction setting $\theta$ and design
$\theta^*_1,\ldots,\theta^*_{37}$. Specifically,
\begin{eqnarray*}
\mu_\theta &=& r(\theta)R(\theta^*;\vartheta)^{-1}w^* \\
\sigma^2_\theta &=&
\sigma^2[1-r(\theta)R(\theta^*;\vartheta)^{-1}r(\theta)'], 
\end{eqnarray*}
where 
\begin{equation}
 r(\theta)=\left\{R(\theta,\theta^*_1;\vartheta),\ldots,
  R(\theta,\theta^*_{37};\vartheta)\right\}.
\end{equation}
The emulator uses the conditional mean $\mu_\theta$ as the
estimate for $w(\theta)$.  The prediction for the power spectrum
${\cal P}(k,z;\theta)$ can then be produced using Eqn.~(\ref{model}).

\subsubsection{Emulator Performance}

\begin{table}
\begin{center} 
\caption{\label{tab:test} Parameters for 10 Test Models}
\begin{tabular}{ccccc}
\tableline\tableline
$\omega_m$ & $\omega_b$ & $n_s$ & $-w$ & $\sigma_8$ \\
\hline 
0.1327 & 0.0220 & 0.8890 & 0.8406 & 0.7235 \\
0.1524 & 0.0219 & 0.8792 & 0.9515 & 0.7926 \\
0.1542 & 0.0224 & 0.8533 & 1.1044 & 0.8630 \\
0.1428 & 0.0235 & 1.0302 & 1.1359 & 0.8740 \\
0.1372 & 0.0226 & 0.9474 & 0.7219 & 0.8877 \\
0.1337 & 0.0228 & 0.9499 & 0.7392 & 0.7985 \\
0.1386 & 0.0229 & 1.0260 & 0.8145 & 0.7870 \\
0.1490 & 0.0228 & 0.9979 & 1.1144 & 0.8156 \\
0.1250 & 0.0229 & 0.9031 & 0.7963 & 0.7916 \\
0.1214 & 0.0228 & 1.0043 & 0.9418 & 0.7555 \\
\tableline\tableline
\end{tabular}
\end{center}
\end{table}

\begin{figure}[h]
\centerline{
 \includegraphics[width=4in]{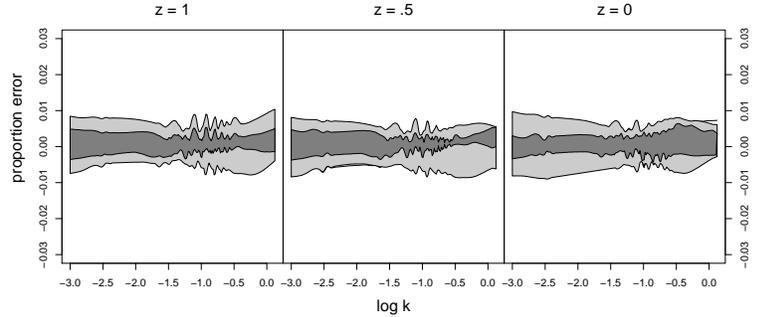}}
\caption{\label{emul}Emulator performance at three redshifts,
  $z=1,~0.5$ and 0 (left to right). The emulator is tested on 10
  additional {\sc HaloFit} runs within the parameter priors. The
  emulator error with respect to the {\sc HaloFit} results is
  shown. The central gray region contains the middle 50\% of the
  residuals, the wider light gray region, the middle 90\%. Errors are
  at the sub-percent level. We emphasize that we do not show the
  average of the residuals, which would artificially suppress the
  error, but rather represent the residuals by the two bands.}
\end{figure}

In order to evaluate the accuracy of the emulator we generate a second
set of ten power spectra with {\sc HaloFit} within the prior parameter
ranges for three redshifts, $z=0,~0.5,$ and 1. For this set we choose
the input cosmologies randomly, but still insuring that the constraint
on the Hubble parameter is satisfied. The ten additional cosmologies
are listed in Table~\ref{tab:test}. We then predict the results for
those cosmologies with the emulator scheme and compare them to the
{\sc HaloFit} output, the ``truth'' in this case. The results for the
residuals are shown in Figure~\ref{emul}. 

The middle 50\% of the residuals (dark gray band) are accurate to
0.5\% or better over the entire $k$-range and for all three
redshifts. All predictions have errors less than 1\%. This result
shows that a simulation set with as small a number as 37 cosmologies
is sufficient to produce a power spectrum emulator accurate at 1\%.

In \citet{HHHNW} several other convergence tests are discussed and
demonstrated. These tests show that emulator performance can improve
considerably -- by an order of magnitude -- if either the number of
simulation training runs is increased or the parameter space under
consideration is narrowed. In the present paper we follow the second
strategy, restricting the priors as much as is sensible given the
current and near-future observational situation.

\subsection{Sensitivity Analysis}

\begin{figure}[b]
\centerline{
 \includegraphics[width=3.5in]{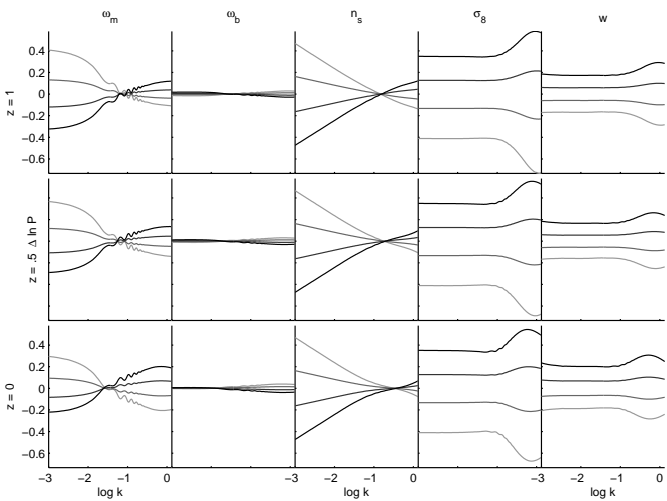}}
\caption{\label{sens} Sensitivity analysis for each of the five
  parameters at redshift $z=0,~0.5,$ and 1 (bottom to top).  The
  $y$-axis shows the deviation of the log of the power spectrum from
  the nominal spectrum where each parameter is set at its
  midpoint. The light to dark lines correspond to the smallest
  parameter setting to the largest. Due to the tight constraints on
  $\omega_b$ from CMB measurements, which led us to choose a rather
  narrow prior, $\omega_b$ variation leads to very little change in
  the power spectrum. Creating this plot from simulations directly
  would have required additional 20 simulations, costing 
  $\sim$400,000 CPU-hours. }
\end{figure}

After the emulator has been built it can be used to explore the
behavior of the power spectrum -- within the parameter priors -- in more 
detail than if one had access only to the results at the design
points. Because this can be done quickly and in a straightforward
manner, an obvious use of emulation is to
carry out a sensitivity analysis, i.e. study the behavior of the power
spectrum as the underlying cosmological input parameters are varied.
It is also very useful to carry out a sensitivity analysis of
quantities related to the GP modeling itself, such as the principal
component weight functions and the correlation functions. Such an
analysis is important to ensure that the GP model is robust and
accurate and will lead, in addition, to more insight about the global
importance of different cosmological parameters.

A first example of a sensitivity analysis is represented in
Figure~\ref{sens}. Here we show variations of the power spectrum at
three redshifts $z=0$, $z=0.5$, and $z=1$. The reference power
spectrum is that obtained with every parameter fixed at the midpoint
of its prior range, i.e., in this case, for the cosmology
$\omega_m=0.1375$, $\omega_b=0.02215$, $n_s=0.95$, $w=-1$,
$\sigma_8=0.758$. (This power spectrum is close to the mean of the 37
models from our design, but not the same.) Next, only one parameter is
varied between its maximum and minimum value while the other four
parameters are fixed at their midpoints.  In Figure~\ref{sens} from
left to right we vary $\omega_m$, $\omega_b$, $n_s$, $\sigma_8$, and
$w$, showing the difference between natural logarithms of these two
power spectra. We note that the Hubble parameter is different for each
power spectrum shown in the figure since it is separately optimized
for each cosmology.

The results contain information about the scales at which the power
spectrum is most sensitive to each parameter and about parameter
degeneracies.  For example, it is clear that the power spectrum is
relatively insensitive to $\omega_b$, within the allowed range, at any
scale or redshift and it will therefore be difficult to further
constrain $\omega_b$ from power spectrum measurements alone.  In the
quasi-linear to nonlinear regime at $k\sim 0.1-1\,h\,{\rm Mpc}^{-1}$,
the power spectrum holds significant information regarding $\sigma_8$
and $w$, but degeneracies become an issue.  Very large scales are
particularly sensitive to the spectral index and $\omega_m$, which
sets the epoch of matter-radiation equality and hence the position of
the peak in the power spectrum.

\begin{figure*}
\centerline{
 \includegraphics[width=5.5in]{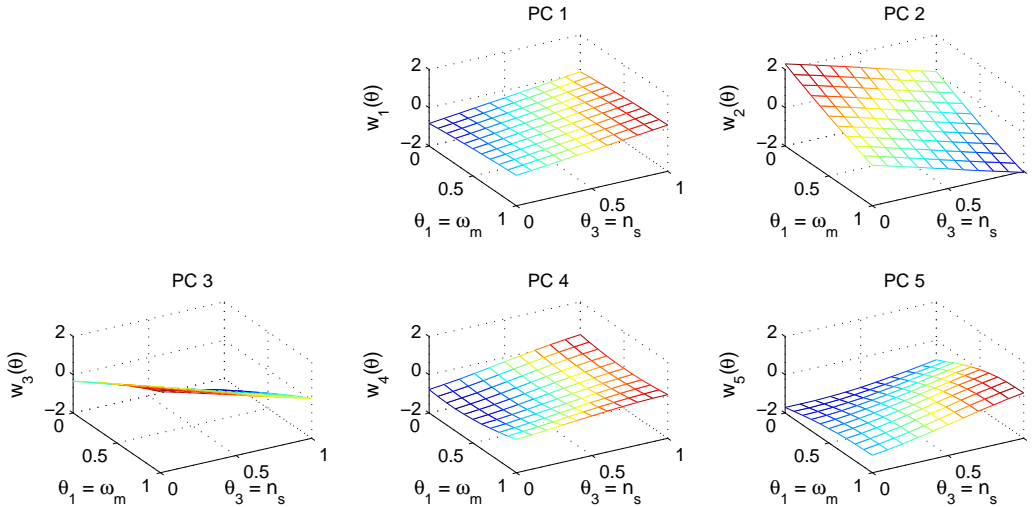}}
\caption{\label{wsurf}Posterior mean estimates of the
  principal component weight functions $w_1(\theta)$ to $w_5(\theta)$.
  Here the prediction points $\theta$ are over a grid of 
  $\omega_m$ and $n_s$ values, while the remaining parameters are
  held fixed at their midpoints. The
  cosmological parameters are displayed in the normalized [0,1]
  space.}
\end{figure*}

Next we investigate the change of the PC weight functions under the
influence of varying cosmological parameters.  Figure~\ref{wsurf}
shows the results for the PC weights $w_1(\theta)$ to $w_5(\theta)$
corresponding to the PCs shown in Figure \ref{basis}.  We show the
results as a function of two of the active parameters, $\omega_m$ and
$n_s$, while holding the remaining three parameters fixed at their
midpoints.  Note the very smooth behavior of the weights as a function
of the parameters.  The behavior of the PC weight functions can be
loosely connected to the correlation functions in the
following way: if the posterior mean surface for the PC weight
functions looks well behaved (no steep or abrupt changes) then the
correlation functions are likely away from zero.  This means the surface is easy to
predict at untried input settings.  However, if it changes
drastically, one or more of the correlation functions is near zero, and the GP will
not give very accurate predictions for $w_i(\theta)$ at new input
settings, as discussed above.

The sensitivity analysis is also helpful in the targeted augmentation
of simulation designs. If the accuracy of the emulator is not
sufficient for the problem of interest, one would like to improve it
by adding additional simulations. These simulations would then involve
variations of the most active parameters while keeping the other
parameters more or less fixed. The augmentation of existing designs is
an active field of research in statistics, and potentially important
in precision cosmology applications.

\section{Conclusions}
\label{conclusion}

The last three decades have witnessed unprecedented progress in
cosmology. From order of magnitude and factor of two estimates for
cosmological parameters, we now have measurements at 10\% accuracy or
better. These measurements have revealed one of the biggest mysteries
in physics today: a dark energy leading to the acceleration of the
expansion of the Universe. In order to understand the origin, nature,
and dynamics of this dark energy -- or to prove that the acceleration
is due to a modification of gravity on the largest length scales --
the accuracy of the measurements must be further improved. The next
step, as defined by near-term and next-generation surveys is to obtain
measurements at the 1\% accuracy level. This puts considerable stress
on the quality of theoretical predictions, which have to be at least
as accurate. Four major probes of dark energy -- baryon acoustic
oscillations, weak lensing, redshift space distortions and clusters --
are based on measurements of the large scale structure in the Universe.
In order to obtain precise predictions for these probes, expensive,
nonlinear simulations have to be carried out and ways must be found to
extract the needed information from a limited number of such simulations.

In this paper, we have demonstrated that if very accurate simulations
are available, 1\% accurate prediction schemes can be produced from
just tens of high-accuracy simulations. The focus of this paper is the
nonlinear matter fluctuation power spectrum, but the general scheme
applies to any other cosmological statistic, e.g., the halo mass
function, statistics of extrema, higher order functions, velocities,
etc.

In \citet{Heitmann08} we introduced a set of 38 cosmological
simulations, the Coyote Universe suite, all of which satisfy the 1\%
error control criterion for the power spectrum up to $k\sim 1\,h\,{\rm
  Mpc}^{-1}$.  In the current paper we demonstrate (on {\sc HaloFit\/}
generated spectra) that from these simulations we can generate an
emulator for the nonlinear power spectrum which has essentially the
same accuracy as the simulations themselves.  The high accuracy
attained from a small number of simulation inputs is due to (i)
understanding the physical variables to use as inputs to the
interpolation {}from our detailed knowledge of cosmological
perturbation theory, (ii) an interpolation method based on a
sophisticated simulation design and GP modeling which has been
developed and refined in the statistics community over the last decade
to address problems of the nature described here, and (iii) the
excellent parameter constraints from CMB measurements, which allow us
to base our emulator on relatively narrow parameter priors and
therefore ease the interpolation task.

This paper is the second in a series of three papers with the final
goal to provide a high-precision emulator for the nonlinear power
spectrum out to $k\sim 1\,h\,{\rm Mpc}^{-1}$.  The first paper of the
series \citep{Heitmann08} demonstrated that the matter power spectrum
could be calculated to $\mathcal{O}(1\%)$ from well controlled
$N$-body simulations.  The current paper introduces the cosmologies
underlying the Coyote Universe simulation suite, explaining and
demonstrating success of the emulation technology using {\sc
  HaloFit\/} as a proxy for the simulation results.  Our prediction
scheme can achieve 1\% accuracy from only a limited number of
simulations: approximately 37 cosmological models are adequate for
this purpose.  The third and final paper will present results from the
simulation suite discussed in this paper and will include a power
spectrum emulator built around them.  This emulator will be publicly
released.

\acknowledgements

Part of this research was supported by the DOE under contract
W-7405-ENG-36, and by a DOE HEP Dark Energy R\&D award. SH, KH, DH,
and CW acknowledge support from the LDRD program at Los Alamos
National Laboratory.  MJW was supported in part by NASA and the DOE.
We would like to thank Dragan Huterer, Lloyd Knox, Nikhil Padmanabhan,
and Michael Schneider for useful discussions. K.H. is grateful to the
SAMSI Summer School on the Design and Analysis of Computer Experiments
for hospitality and especially thanks Jerry Sacks and Will Welch for
their outstanding lectures, which guided some of the introductory
section on GP modeling in this paper.

\appendix

\section{Optimization of space filling designs}
\label{appopt}

\begin{figure}[t]
\centerline{
 \includegraphics[width=3.3in]{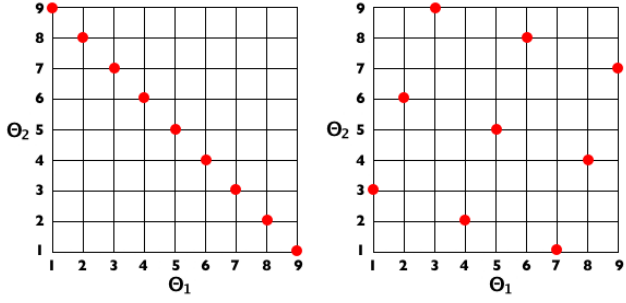}}
\caption{\label{design_app}Two random LH designs. The left panel is
  clearly an unfortunate design, not fulfilling our requirements about
  space-filling. The right panel shows an LH design in which the
  minimum distance between points is maximized. The space-filling
  properties of this design are much better.}
\end{figure}

As mentioned in Section~\ref{sec:design} none of the discussed design
strategies (OA, LH, OA-LH, SLH) provide a unique ``best'' design.  For
a given number of parameters, $m$, and the number of simulations, $n$,
a large number of possible designs exist (e.g.~for LH designs, the
number is $(n!)^{m-1}$).  The question is then how to choose the most
suitable design for a given problem.  A major requirement for the
designs used in this paper is that they should have good space-filling
properties.  Figure~\ref{design_app} shows two LH designs.  As
explained in Section \ref{sec:design}, in an LH design in every column
every entry appears only once, which is clearly fulfilled for both
designs shown.  In the panel on the right the minimum distance between
points was maximized, clearly leading to better space-filling
properties.  In the following we will describe two possible
optimization criteria for OA-LH designs and for SLH designs which have
been used in our work.  We closely follow the discussion in
\cite{santner03}.

\subsection{Maximin Distance Design}
\label{sec:maxmin}

Our aim in optimizing the space-filling properties in our design is
that no two points in the design are too close. In the example in
Figure~\ref{design_app} we use the maximin distance optimization to
spread the points out in the two dimensional plane.  This optimization
scheme maximizes the minimum distance between points. A more general
approach is to minimize the ``average'' of some function of the
distances between pairs of design points. As a first step, we have to
define a distance measure between design points. Suppose we have an
arbitrary $n$-point design $\cal D$ with input settings $\{{\bf
x}_1,{\bf x}_2,\dots, {\bf x}_n\}$ and we want to vary $m$ parameters.
The $p^{\rm th}$ order distance between two input settings is defined
as
\begin{equation}
  \rho_p({\bf x},{\bf x}')=
  \left[\sum_{j=1}^m \left|x_j-x'_j\right|^p\right]^{1/p}
  \quad ,\quad p\ge 1.
\label{dist}
\end{equation}
For $p=1$ this is known as the rectangular distance, for $p=2$ it is
the Euclidean distance. Next we define the average distance criterion
as follows: 
\begin{equation}
  d_{(p,\lambda)}({\cal D})=\left(\frac{2}{n(n-1)}
  \sum_{{\bf x}_i,{\bf x}_j}
  \left[\frac{m^{1/p}}{\rho_p({\bf x}_i,{\bf x}_j)}
  \right]^\lambda\right)^{1/\lambda}
  \quad , \quad \lambda\ge 1.
\label{crit}
\end{equation}
The combinatorial factor $n(n-1)/2$ is simply the number of different
pairs, $({\bf x}_i,{\bf x}_j)$, that can be drawn from the $n$ points
in the design ${\cal D}$ and $m^{1/p}$ is the maximum distance between
two points in $[0,1]^m$ (where we have normalized the domain of each
input variable to $[0,1]$ as in the main paper):
\begin{equation}
  0 < \rho_p({\bf x_i,\bf x_j})\le m^{1/p}.
\end{equation}
For fixed $(p,\lambda)$, an $n$-point design ${\cal D}_{av}$ is optimal if
\begin{equation}
  d_{(p,\lambda)}({\cal D}_{av})={\rm min}~d_{(p,\lambda)}({\cal D}).
\label{max_d}
\end{equation}
For example, if $\lambda=1$, this condition will lead to a design
which avoids ``clumpiness''.  The optimal average distance designs
might not be optimal in projected spaces.  In the main paper we
mentioned that often only a few parameters are active (for example for
the matter power spectrum the main active parameters are $\sigma_8$
and $\omega_m$) and therefore it is desirable for designs to have good
coverage if projected down onto two or three dimensions.  Such designs
can be found by computing Eq.~(\ref{crit}) for each relevant
projection of the full design ${\cal D}$ and averaging these to form a
new function which is then minimized.  An implementation of this
approach can be found in \cite{welch85}.  To be more concrete, let us
start again with a candidate design ${\cal D}$.  Now we want to
consider the projection in $j$ dimensions.  This will lead to designs
${\cal D}_{kj}$ which are the $k$-th projection onto $j$ dimensions.
Following our example in the main text, where we discussed a design
with $m=3$ parameters and we want good space-filling properties onto
$j=2$ dimensions (as shown in Figure \ref{design_2d}), we have $k=3$
projections.  Following Eq.~(\ref{crit}), the average distance
criterion function for the projected design ${\cal D}_{kj}$ is given
by
\begin{equation}
  d_{(p,\lambda)}({\cal D}_{kj})=\left(
  \frac{2}{n(n-1)}\sum_{{\bf x}^*_h,{\bf x}^*_i}
  \left[\frac{j^{1/p}}{\rho_p({\bf x}^*_h,{\bf x}^*_i)}
  \right]^\lambda\right)^{1/\lambda}
\end{equation}
where ${\bf x^*}_l$ is the projection of ${\bf x}_l$ onto ${\cal D}_{kj}$.
We can now define the average projection design criterion function to be:
\begin{equation}\label{crit_av}
  {\rm av}_{(p,\lambda)}({\cal D})=
  \left(\frac{1}{\sum_{j\in J}C(m,j)}\sum_{j\in J}\sum_{k=1}^{C(m,j)}
\left[d_{(\rho,\lambda)}({\cal D}_{kj})\right]^\lambda\right),
\end{equation}
with $C(m,j)=m!/[j!(m-j)!]$.
An $n$-point design, ${\cal D}_{avp}$, is optimal with respect to
the criterion given in Eq.~(\ref{crit_av}) if
\begin{equation}
  {\rm av}_{(p,\lambda)}({\cal D}_{avp})=
  {\rm min}\ {\rm av}_{(p,\lambda)}({\cal D}).
\end{equation}
Now that we have defined the average projection design criterion
function, including good projection properties, we can judge if a
design is close to optimal.  For the design used in the main part of
the paper, we used two different optimization algorithms and picked
the optimal design in the end by choosing the one with the better
performance with respect to the distance based criterion.  In the
following, we will briefly outline the two algorithms.  Both
algorithms were used to optimize symmetric LH designs.

\subsection{Simulated Annealing Algorithm for Optimized SLH Designs}
\label{SA}

The simulated annealing (SA) algorithm was introduced by \cite{mm95}
to search for optimal LH designs. We adopt the algorithm here for
symmetric and non-symmetric LH designs. The basic idea is to exchange
points in the design within a column, evaluate the quality of the new
design by a distance criterion, and keep the new design if it is
better than the previous. To be more concrete, the algorithm begins
with a randomly chosen LH design for which the distance criterion is
measured. Then a random column is picked and within that column two
randomly picked elements are exchanged. If the design is symmetric, it
is important to do the exchange in sets of pairs to keep the symmetry
of the design. Suppose in a 4-row SLH design element 1 is exchanged
with element i, then element 4 must be exchanged with element
$5+1-i$. No exchange is needed if element $i$ is exchanged with
element $n+1-i$ (for an excellent description of the procedure see
\citealt{ye00}). If the design has an odd number of rows, the center
point -- which must be a design point -- remains untouched.  In this
way, a new design ${\cal D}_{\rm try}$ is generated. The quality of
the new design is measured via the distance criterion and if the
design is better than the previous one it is kept. If the design is
worse, it will replace ${\cal D}$ with the probability
$\pi=\exp\{-[\phi({\cal D}_{\rm try})-\phi({\cal D})]/t\}$ where $t$
is a preset parameter, referred to as temperature (the form of the
probability gave the algorithm its name ).  $\phi({\cal D})$ is the
criterion value for the design ${\cal D}$. This procedure works like a
Markov Chain Monte Carlo, the idea being that after some time the
optimization procedure will reach a local minimum close to an optimal
design.  The value for the temperature $t$ influences the search area
for the optimal design, the higher the value of $t$, the more global
the search will be.  Obviously, this will slow down the search, since
designs which might not be very good, are kept with higher
probability.  In practice, the algorithm is stopped after some time
and restarted from a range of random initial designs. In the end, the
design with the best value for the distance criterion is kept.

\begin{figure}[t]
\centerline{
 \includegraphics[width=3.3in]{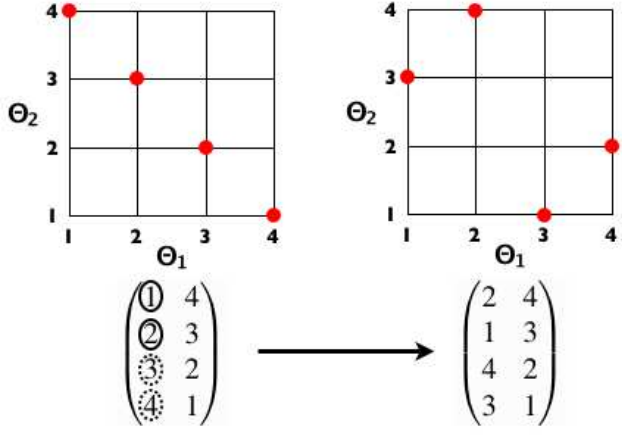}}
\caption{\label{fig:SA}A random symmetric LH design with two
  parameters ($m=2$) and four design points ($n=4$). The first two
  elements that are exchanged are marked by the circle, the pair that
  has to be exchanged in addition to keep the symmetry is marked by
  dashed circles.  The right panel shows an improved design after one
  SA step. The value for the distance criterion for the second design
  signals improvement, points are spread out more evenly.}
\end{figure}

A simple example for the first step in an SA algorithm is shown in
Figure~\ref{fig:SA}. We start with a ``random'' design, in this case
all design points are on the diagonal. Note, that this is an
acceptable symmetric LH design. In this example, we have $n=4$ design
points and $m=2$ parameters.  This design is obviously not optimal,
the points in the design are not evenly spread out and there are many
small distances between the points close to the center of the
plane. Following Eq.~(\ref{dist}), choosing $p=2$ for an Euclidian
distance measure and normalizing all design points to $[0,1]^2$, we
find that the maximum distance is $\sqrt{2}$ (between the two corner
points), obeying Eq.~(\ref{max_d}). The value for the distance
criterion given by Eq.~(\ref{crit}) is easily calculated (we choose
$\lambda=1$): $d_{(2,1)}({\cal D})\simeq 2.2$. Now we try to improve
the design following the scheme outlined above. We pick randomly
column 1 and exchange the first two elements. To keep the symmetry, we
have to also exchange element 3 and 4. The new design has a better
value for the distance criterion, $d_{(2,1)}({\cal D_{\rm try}})\simeq
1.95$, we therefore keep the design. Now we can repeat the procedure
until we find an optimal design or we decide to stop the process. The
first new design is shown in the right panel in Figure~\ref{fig:SA}
and is clearly better than the original design.

\subsection{Columnwise-pairwise Algorithm for Optimized SLH Designs}
\label{CP}
As for the SA algorithm, the columnwise-pairwise (CP) algorithm is
based on columnwise-pairwise exchanges. The first major difference
between the two algorithms is that the CP algorithm stops once a new
design is found that is better than the previous. The second major
differences is that, for each column, elements are exchanged until the
best configuration for each column is found. In this regard, the CP
algorithm is a more local algorithm and will most likely converge to a
local optimum and rarely to a global optimum.  Following \cite{ye00},
the algorithm can be summarized as follows:
\begin{enumerate}
\item As in the case of an SA algorithm, start with a random SLH design.
\item Each iteration has $m$ steps, one for each column.  At the $i\,$th
step, the best two simultaneous exchanges within column $i$ are found
(remember, in order to keep the symmetry, one always has to do two
simultaneous exchanges).  The term ``best exchanges'' refers again to the
quality of the global design with respect to a distance criterion.
The design matrix is updated accordingly.
\item If the overall new design is better with respect to the distance
criterion, repeat Step 2.  Otherwise, the new design is considered to be
``optimal'' and the search is terminated.
\end{enumerate}
As for the SA algorithm one should create several optimal designs starting
{}from different initial designs and determine which one of the resulting
designs is the best one.

\section{Gaussian Process Modeling of the PC weight functions}
\label{app_gp}

\begin{table*}
\begin{center} 
\caption{\label{tab:gp} Definitions and Explanations of Variables Used
in this Section}
\begin{tabular}{ll}
\tableline\tableline
Variable number/value & Explanation \\
\hline 
$\Phi=[\phi_1;...\phi_5]$  & PC basis matrix, consisting of orthogonal
basis vectors\\ 
${\cal P}=\ln [\Delta^2(k,z)/(2\pi k^{3/2})]$ & Rescaled power spectrum
\\ 
$n_{\cal P}$=5 & Number of principal components \\
$n_\theta=5$ & Number of cosmological parameters \\
$\theta=\omega_m,~\omega_b,~n_s~,w~,\sigma_8$ & Cosmological
parameters\\
$w_i(\theta)$ & GP model of PC weights for $i^{\rm th}$ basis
function\\ 
$w^*_i$ &  GP model of PC weights for $i^{\rm th}$ basis function at
the design settings $\theta^*_1,\ldots,\theta^*_m$\\
 $w^*$ & GP model of all PC weights at design inputs:
 vec$(w^*_1,\ldots,w^*_{n_{\cal P}})$\\ 
$\hat w$ & PC weights obtained from projecting simulation output onto
bases; contains numerical error \\ 
$m=37$ & Number of cosmological models\\
$\lambda_{w_i}$ & Marginal precision of the GP process for the $i^{\rm
  th}$ PC\\ 
 $\rho_w$ & Correlation functions \\ 
$\lambda_{\cal P}$ & Precision describing simulation and truncation
error\\ 
$n_{kz}=20000$ & Number of points for each power spectrum (200
$k$-values at 100 redshifts)\\
$a_{\rho_w}=1,~b_{\rho_w}=0.1$ & Prior parameters in the
$\beta$-distribution for  the $\rho_{w;il}$'s\\ 
$a_{w}=5,~b_{w}=5$ & Prior parameters in the $\Gamma$-distribution for
the $\lambda_{w_i}$'s\\ 
\tableline\tableline
\end{tabular}
\end{center}
\end{table*}

In this appendix we discuss the full GP modeling process for the PC
weights. The major differences between the discussion here and the
discussion in the main text are:

\begin{itemize}
\item The dimensionality of the problem is increased.
\item A full Baysian treatment of the problem is performed, including
full priors for correlation parameters.
\item We drop the assumption that our measurements are perfect and
introduce  $\lambda_{\cal P}$ (see below) to account for simulation
errors and errors due to the truncation in the basis functions. 
\item We use a slightly different correlation function.
\end{itemize}

Consider the full problem stated in the text, where $\theta$ lives in
a $n_{\cal P}=5$ dimensional space and represents a cosmology with
$n_\theta=5$ input parameters.  We model each PC weight function
$w_i(\theta)$, $i=1,\ldots,5$ as a mean-zero GP
\begin{equation}
  w_i(\theta)\sim GP(0,\lambda_{w_i}^{-1}R(\theta,\theta';\rho_{w_i})), 
\label{gpmodel}
\end{equation}
where the symbol $\sim$ means ``distributed according to''.
Here $\lambda_{w_i}$ is the marginal precision of the process and the
correlation function is given by:
\begin{equation} \label{corApp}
  R(\theta,\theta';\rho_{w_i})= \prod_{l=1}^{n_\theta}
  \rho_{w;il}^{4(\theta_l-\theta_l')^2}. 
\end{equation}
This form is mathematically equivalent to that of Eq.~(\ref{sigma}) --
set $n_\theta = 1$, $i=1$, and $\rho=e^{-\vartheta/4}$.  The parameter
$\rho_{w;il}$ controls the spatial range for the $l$th input dimension
of the process $w_i(\theta)$.  Under this parametrization,
$\rho_{w;il}$ gives the correlation between $w_i(\theta)$ and
$w_i(\theta')$ when the input conditions $\theta$ and $\theta'$ are
identical, except for a difference of 0.5 in the $l$th component.  Our
task is now to find $\lambda_{w_i}$ and $\rho_{w;il}$ from the set of
our simulations.

From our 37 simulations, we first define a 5-component, 37-vector
$w_i$ with $i=1,...,5$:
\begin{equation}
  w^*_i=(w_i(\theta^*_1),...,w_i(\theta_{37}^*))^T.
\end{equation}
The star indicates that we use our 37 input cosmologies here and
therefore the answer for ${\cal P}$ is known at that point.  Assume
that $w^*$ is normal-distributed with mean zero:
\begin{equation}
  w^*\sim N(0,\Sigma_w),
\end{equation}
where $\Sigma_w={\rm diag}(\Sigma_{w_1},\ldots,\Sigma_{w_5})$ and
$\Sigma_{w_i}\equiv\lambda^{-1}_{w_i}R(\theta^*;\rho_{w_i})$ -- the
$37 \times 37$ matrix obtained by applying Eq.~(\ref{corApp}) to each
pairwise combination of the 37 design points
$\theta^*_1,\ldots,\theta_{37}^*$.  $\Sigma_w$ is therefore controlled
by five precision parameters $\lambda_w$ and the 25 spatial
correlation parameters held in $\rho_w$.  Next we have to specify
priors for each $\lambda_{w_i}$ and for the $\rho_{w;il}$.  Following
\citet{HHHNW}, we choose $\Gamma(a_w,b_w)$ distributions for the
priors for $\lambda_{w_i}$ and $\beta(a_{\rho_w}, b_{\rho_w})$ priors
for the $\rho_{w;il}$:
\begin{eqnarray}
\pi(\lambda_{w_i})&=&\frac{1}{\Gamma(a_w)}
b_w(b_w\lambda_{w_i})^{a_w-1}e^{-b_w\lambda_{w_i}},
\quad i=1,...,5,\\
\pi(\rho_{w;il})&=&\frac{\Gamma(a_{\rho_w}+b_{\rho_w})}
{\Gamma(a_{\rho_w})\Gamma(b_{\rho_w})}\rho_{w;il}^{a_{\rho_w}-1}\nonumber\\
&&\times (1-\rho_{w;il})^{b_{\rho_w}-1}
\ i=1,\ldots,5,~~~l=1,\ldots,5
\end{eqnarray}
with $a_w=b_w=5$, $a_{\rho_w}=1$, and $b_{\rho_w}=0.1$. The choices for
$a_w$ and $b_w$ lead to a prior for $\lambda_{w_i}$ of mean 1 and a
prior standard deviation of 0.45 (The mean of a $\Gamma$ distribution
is given by $a/b$ and the standard deviation by $\sqrt{a/b^2}$). The
choice of unit mean is consistent with the standardization of the GP
for the $w_i$.

The choices for $a_{\rho_w}$ and $b_{\rho_w}$ lead to a substantial
prior mass near 1 [The mean of a $\beta$ distribution is given by
$a/(a+b)$ and the standard deviation by $\sqrt{ab/((a+b)^2(a+b+1))}$.]
In general, the selection of these parameters depends on how many of
the $n_\theta$ inputs are expected to be active.

Now we return to the {\em actual} information that we have, and from
which we want to derive the weights $w_i$: the simulation outputs for
the power spectra ${\cal P}^*$ for the 37 cosmologies. We arrange
these outputs in an $n_{kz} m$ vector
\begin{equation}
{\cal P}^*={\rm vec}([{\cal P}(\theta_1^*);\cdots;{\cal
  P}(\theta_{37}^*)]). 
\end{equation}
The simulation outputs have two sources of error: the error intrinsic
to the simulation (e.g., realization scatter, numerical error) and the
error due to the truncation in basis functions used to model ${\cal
  P}(k;\theta)$ via Eq.~(\ref{model}). We encapsulate the precision of
the error in $\lambda_{\cal P}$ and we assume that the error
$\epsilon$ itself in Eq.~(\ref{model}) is independent and identically
normal distributed. We are now in a position to formulate the
likelihood for ${\cal P}^*$:
\begin{equation}\label{likeli}
p({\cal P}^*|w^*,\lambda_{\cal P})\propto\lambda_{\cal P}^{m n_{kz}/2} 
\exp\{-\frac{1}{2}\lambda_{\cal P}({\cal P}^*-\Phi w^*)^T({\cal P}^*-\Phi w^*)\},
\end{equation}
where $\Phi$ is a matrix composed from the $\phi_i$ basis vectors
which we use to model the power spectra (see Eq.~(\ref{model})). As
for $\lambda_w$, we specify the priors by a $\Gamma (a_{\cal
  P},b_{\cal P})$ distribution. We expect the data to be very
informative about $\lambda_{\cal P}$ and therefore choose the prior to
be very broad with $a_{\cal P}=1$ and $b_{\cal P}=0.0001$. This prior
allows for large values of $\lambda_{\cal P}$ which force the GP model
to nearly interpolate the simulation output.  This will happen when
the PC representation of the output is very good.

This result is only an intermediate step, as our goal is to find the
likelihood for the $w_i$ not for ${\cal P}$ itself. Fortunately, we
can factorize Eq.~(\ref{likeli}) to extract the likelihood for the
weights easily. To do this we define $\hat w$ as
\begin{equation}
\hat w=(\Phi^T\Phi)^{-1}\Phi^T{\cal P}^*.
\end{equation}
Note that $\hat w$ encapsulates the error due to the truncation of the
basis functions in modeling ${\cal P}^*$. With this definition, it is
easy to show that Eq.~(\ref{likeli}) can be written as
\begin{eqnarray}\label{likeli2}
p({\cal P}^*|w^*,\lambda_{\cal P})&\propto& \lambda^{m n_{\cal P}/2}_{\cal P}
\exp\{-\frac{1}{2}\lambda_{\cal P} (w^*-\hat w)^T(\Phi^T\Phi)
(w^*-\hat w)\}\nonumber\\
&&\times \lambda_{\cal P}^{m(n_{kz}-n_{\cal P})/2}\\
&&\times\exp\{-\frac{1}{2}\lambda_{\cal P}{\cal P}^{*T}
(I-\Phi(\Phi^T\Phi)^{-1}\Phi^T){\cal P}^*\}\nonumber,
\end{eqnarray}
with $m=37$, $n_{\cal P}=5$, and $n_{kz}$ denoting the $(k,z)$ support
for each power spectrum is measured. Note that in the first line of
Eq.~(\ref{likeli2}), $\hat w$ is completely separated from the rest of
the likelihood expression. We can use this factorization to represent
the likelihood in a dimension-reduced form:

\begin{equation}
p(\hat w|w^*,\lambda_{\cal P})\propto
 \lambda^{m n_{\cal P}/2}_{\cal P}
\exp\{-\frac{1}{2}\lambda_{\cal P} (w^*-\hat w)^T(\Phi^T\Phi)(w^*-\hat w)\},
\end{equation}
where the remaining terms from Eq.~(\ref{likeli2}) are absorbed in a
re-defined Gamma distribution prior for $\lambda_{\cal P}$,
$\Gamma(a'_{\cal P},b'_{\cal P})$ with 
\begin{eqnarray}
a'_{\cal P}&=&a_{\cal P}+\frac{m(n_{kz}-n_{\cal P})}{2},\\
b'_{\cal P}&=&b_{\cal P}+\frac{1}{2}{\cal P}^T
(I-\Phi(\Phi^T\Phi)^{-1}\Phi^T){\cal P}.
\end{eqnarray}

It is useful to recap what has been done so far: We began with the
normal likelihood for ${\cal P}$ with the Gamma distribution prior for
$\lambda_{\cal P}$:
\begin{equation}
{\cal P}^* |w^*,\lambda_{\cal P} \sim N(\Phi
w^*,\lambda_{\cal P}^{-1} I_{n_{kz}}),
~~~\lambda_{\cal P} \sim \Gamma(a_{\cal P},b_{\cal P}).
\end{equation}
Due to the relation
\begin{eqnarray}
p({\cal P}^* |w^*,\lambda_{\cal P})\times \pi(\lambda_{\cal P};a_{\cal P},b_{\cal P})
&\propto& p(\hat w | w,\lambda_{\cal P})\\
&&\times \pi(\lambda_{\cal P};,a'_{\cal P},b'_{\cal P}),\nonumber
\end{eqnarray}
we can derive the likelihood for $\hat w$:
\begin{equation}
\hat w |w^*,\lambda_{\cal P}\sim N(w^*,(\lambda_{\cal P}\Phi^T\Phi)^{-1}),
~~~\lambda_{\cal P}\sim\Gamma(a'_{\cal P},b'_{\cal P}).
\end{equation}
Next, $w^*$ is integrated out, leading to the posterior distribution
\begin{eqnarray}
\label{eq:postw}
\lefteqn{ \pi(\lambda_{\cal P},\lambda_w,\rho_w| \hat{w})  \propto }\\ \nonumber
  & & \left| (\lambda_{\cal P} \Phi^T\Phi)^{-1} + \Sigma_w \right|^{-\frac{1}{2}}\\
  &&\times \exp\{ -\frac{1}{2} \hat{w}^T ([\lambda_{\cal P} \Phi^T\Phi]^{-1} 
+ \Sigma_w)^{-1} \hat{w} \} \\
\nonumber
  & &\times
  \lambda_{\cal P}^{a'_{\cal P}-1} e^{-b'_{\cal P} \lambda_{\cal P}}
  \prod_{i=1}^{n_{\cal P}} \lambda_{w_i}^{a_w-1} e^{-b_w \lambda_{w_i}}
  \prod_{i=1}^{n_{\cal P}}
       \prod_{j=1}^{n_\theta} (1-\rho_{w;ij})^{b_\rho-1}.
\end{eqnarray}
As detailed in \citet{HHHNW} this posterior distribution is a milepost
on the way to creating the emulator for the power spectrum. It can be
explored via MCMC and contains much useful information about the
parametric dependence of the power spectrum, as derived from the
numerical simulation results at the finite number of design points.

\begin{figure}
\centerline{
 \includegraphics[width=3.5in]{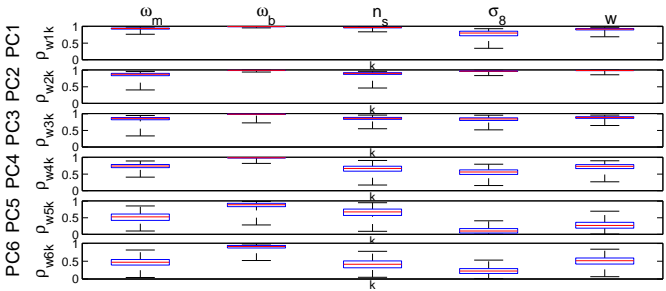}}
\caption{\label{box}Boxplots of posterior samples for each
  $\rho_{w;il}$ for the nonlinear matter power spectrum. The blue box
  itself contains 50\% of the data, the lower edge indicating the 25th
  percentile and the upper edge, the 75th percentile. The red (center)
  line denotes the median. If the red line is not at the center of the
  box, the data is skewed. The black lines (or whiskers) extend out to
  the full range of the data. With our parametrization, a box value
  close to 1 indicates that the parameter is inactive, i.e., the PC is
  not changing much under the variation of that parameter.}
\end{figure}

Next, we briefly discuss the behavior of the correlation function, $\rho_w$,
under the influence of different cosmological parameters.  The results
for the $\rho_w$ are shown in Figure \ref{box} in the form of
boxplots.  Boxplots are commonly used in statistical analyses -- they
offer a convenient way of showing the distribution of data using just
five numbers (see the caption of Figure \ref{box}).

With our definitions \citep[see also][]{HHHNW}), an input $l$ is
inactive for PC $i$ if $\rho_{w;il}=1$.  Inactive here means that the
parameter does not change the actual shape of the power spectrum.  If
$\rho_{w;il}$ is very close to one it can still have strong linear
effects.  In our case, the box values suggest that the coefficients of
the first four PCs are smooth functions of all parameters.  This
implies that it will be easy for the GP model to predict the power
spectrum at untried settings.  If any box value is close to zero, it
indicates there is no smooth functional connection between the
parameters and coefficient values.  In our case, $\sigma_8$ is the
first parameter for which $\rho_w$ is very close to zero for the
coefficient of the fifth principal component, which is the last one we
include in our model.  This will have only a minor impact on
smoothness of the overall input-output relationships as this component
is negligible.  Figure \ref{box} also shows that $\omega_m$ and
$\sigma_8$ are the most active parameters influencing the power
spectrum, as expected on physical grounds.  The equation of state
parameter $w$ is also active, due to the fact that the Hubble
parameter and $\Omega_m$ are changing with $w$.  These observations
are in good agreement with the sensitivity analysis of the power
spectrum itself.

The last step for building the complete emulator is to draw from the
posterior distribution (\ref{eq:postw}) at any given $\theta$. We consider
the joint distribution of $\hat{w}$ and a predicted weight $w_e$ at a new
input parameter setting $\theta_e$:
\begin{equation}
  \begin{pmatrix} \hat{w} \cr w_e \end{pmatrix}
  \sim
  N\left( 0,
  \left[ \begin{pmatrix} (\lambda_{\cal P} \Phi^T\Phi)^{-1} & 0 \cr 0 &
  0 \end{pmatrix} 
  + \Sigma_{w,w_e}(\lambda_w,\rho_w) \right] \right)
\end{equation}
where $\Sigma_{w,w_e}$ is obtained by applying the prior covariance
rule to the augmented input settings that include the original design
and the new input setting $(\theta_e)$. We find
\begin{equation}
\label{eq:emulatorpred}
  w_e|\hat{w} \sim N(V_{21}V_{11}^{-1}\hat{w},
                 V_{22}-V_{21}V_{11}^{-1}V_{12}),
\end{equation}
where
\begin{eqnarray}
  V &=&
  \begin{pmatrix} V_{11} & V_{12} \cr V_{21} &
    V_{22} \end{pmatrix}\nonumber\\
 &=& \left[ \begin{pmatrix} (\lambda_{\cal P} \Phi^T\Phi)^{-1} & 0 \cr 0 &
     0 \end{pmatrix} 
   + \Sigma_{w,w_e}(\lambda_w,\rho_w) \right]
\end{eqnarray}
is a function of the parameters produced by the MCMC output. Hence for
each posterior realization of $\lambda_{\cal P},\lambda_w,\rho_w$, a
realization of $w$ can be produced and the emulator is completed.

\end{document}